\documentclass[twocolumn,amsmath,showkeys,amssymb,superscriptaddress,longbibliography,floatfix, pre]{revtex4-2}

\pdfoutput=1
\usepackage{bm}
\usepackage{graphicx,epsfig}
\usepackage{amsmath,amssymb,mathtools}
\usepackage{url}
\usepackage{hyperref}
\def\be{\begin{equation}}
\def\ee{\end{equation}}

\def\be{\begin{equation}}
\def\ee{\end{equation}}
\def\bc{\begin{center}}
\def\ec{\end{center}}
\def\bea{\begin{eqnarray}}
\def\eea{\end{eqnarray}}

\newcommand{\Avg}[1]{\left\langle{#1}\right\rangle}

\newcommand\mymapsto{\mathrel{\ooalign{$\rightarrow$\cr%
  \kern-.15ex\raise.275ex\hbox{\scalebox{1}[0.522]{$\mid$}}\cr}}}

\def\bea{\begin{eqnarray}}
\def\eea{\end{eqnarray}}

\begin{document}

\title{Designing topological cluster synchronization patterns with the Dirac operator}

\author{Ahmed A. A. Zaid}
\affiliation{School of Mathematical Sciences, Queen Mary University of London, London, E1 4NS, United Kingdom}

\author{Ginestra Bianconi}
\affiliation{School of Mathematical Sciences, Queen Mary University of London, London, E1 4NS, United Kingdom}

\begin{abstract}
Designing stable cluster synchronization patterns is a fundamental challenge in nonlinear dynamics of networks with great relevance to understanding neuronal and brain dynamics. So far, cluster synchronization has been studied exclusively in a node-based dynamical approach, according to which oscillators are associated only with the nodes of the network. Here, we propose a topological synchronization dynamics model based on the use of the Topological Dirac operator, which allows us to design cluster synchronization patterns for topological oscillators associated with both nodes and edges of a network. In particular, by modulating the ground state of the free energy associated with the dynamical model, we construct topological cluster synchronization patterns. These are aligned with the eigenstates of the Topological Dirac Equation that provide a very useful decomposition of the dynamical state of node and edge signals associated with the network. We use linear stability analysis to predict the stability of the topological cluster synchronization patterns and provide numerical evidence of the ability to design several stable topological cluster synchronization states on real connectome data, random graphs, and on stochastic block models.
\end{abstract}

\maketitle

\section{Introduction}
Designing \cite{cho2017stable}, controlling \cite{lehnert2014controlling,tomaselli2025taming}, and engineering \cite{pecora2014cluster,sorrentino2016complete,kiss2007engineering,kori2008synchronization,timofeyev2025cluster} stable cluster synchronization patterns in networks are topics of crucial importance for a wide variety of systems, from chemical oscillators~\cite{kiss2002emerging} to neural and brain networks~\cite{shlens2008synchronized,guevara2017neural}.
Cluster synchronization on simple \cite{sorrentino2007network,ji2013cluster,kim2025cluster} and generalized network structures \cite{jalan2016cluster,salova2021cluster} defines a state in which different clusters of nodes of a network oscillate at different frequencies, often reflecting the underlying symmetries of the network. The current approaches to achieve \cite{cho2017stable,nicosia2013remote} or destabilize \cite{hart2019topological} cluster synchronization rely on symmetries and network fibrations \cite{makse2025symmetries,morone2020fibration,leifer2020circuits}, network topology \cite{lu2010cluster}, time delays, edge weights~\cite{duan2022prevalence}, and phase lags \cite{kiss2007engineering,kori2008synchronization}.

All the approaches proposed so far exclusively address node-based cluster synchronization, i.e., associate the oscillators exclusively with the nodes of the network. In this work, we propose a topological dynamical model, called Dirac-Equation Synchronized Dynamics (DESD) to design patterns of cluster synchronization of topological signals given by oscillators associated with both nodes and edges of the network. We investigate such patterns on real connectomes, random graphs, and block models, and we provide a theoretical framework for predicting their stability. 

 The emerging field of higher-order topological dynamics \cite{millan2025topology,bianconi2021higher} investigates the collective phenomena involving topological signals in networks and simplicial complexes with drastically different properties with respect to the properties of the corresponding node dynamics. This field, combining algebraic topology \cite{bianconi2021higher} and nonlinear dynamics, is transforming our understanding of synchronized dynamical states \cite{millan2020explosive,ghorbanchian2021higher,carletti2023global,wang2024global,arnaudon2022connecting} and higher-order diffusion \cite{torres2020simplicial,ziegler2022balanced,krishnagopal2023topology} and has great potential for brain research \cite{santos2019topological,petri2014homological} and AI \cite{calmon2023dirac,wang2025dirac,nauck2024dirac,battiloro2024generalized}. 
 
 Broadly speaking, topological signals are dynamical variables defined on nodes, edges, and even higher-dimensional simplices of networks and simplicial complexes. Examples of edge signals are ubiquitous and range from synaptic signals and edge signals among brain regions \cite{faskowitz2022edges}, to biological fluxes \cite{barbarossa2020topological} or currents at different locations of the ocean \cite{calmon2022dirac}.

Topological signals are naturally coupled to each other through the Topological Dirac operator \cite{bianconi2021topological}. This operator has its roots in the Kogut and Susskind staggered fermions \cite{kogut1975hamiltonian} defined on lattices and is receiving increasing attention in the study of nonlinear dynamics \cite{calmon2022dirac,calmon2023local,giambagli2022diffusion,muolo2024three,carletti2025global,nurisso2024unified,muolo2024turing} and the development of AI algorithms \cite{calmon2023dirac,wang2025dirac,nauck2024dirac,battiloro2024generalized,alain2023gaussian,wee2023persistent,suwayyid2023persistent}. 

In particular, the Topological Dirac operator leads to the Topological Dirac Equation \cite{bianconi2021topological} that acts on  the topological spinor given by the direct sum of node and edge signals.  
The eigenstates offer a very powerful spectral decomposition of node and edge signals that can be exploited by very advanced AI algorithms~\cite{wang2025dirac,nauck2024dirac}.

In higher-order nonlinear topological dynamics, the Topological Kuramoto model \cite{millan2020explosive,ghorbanchian2021higher} and the Topological Dirac synchronization \cite{calmon2022dirac,calmon2023local}  are fundamental Kuramoto-like dynamical processes that lead to synchronization of topological signals.
The Topological Kuramoto model on a network captures the synchronization of edge signals and it displays a synchronized state aligned along the harmonic eigenvector of the $1$-Hodge Laplacian. The Topological Dirac synchronization \cite{calmon2022dirac,calmon2023local} non-trivially couples node and edge signals associated with the network under study.
\begin{figure*}[!htb!]
     \centering
     \includegraphics[width=1.8\columnwidth]{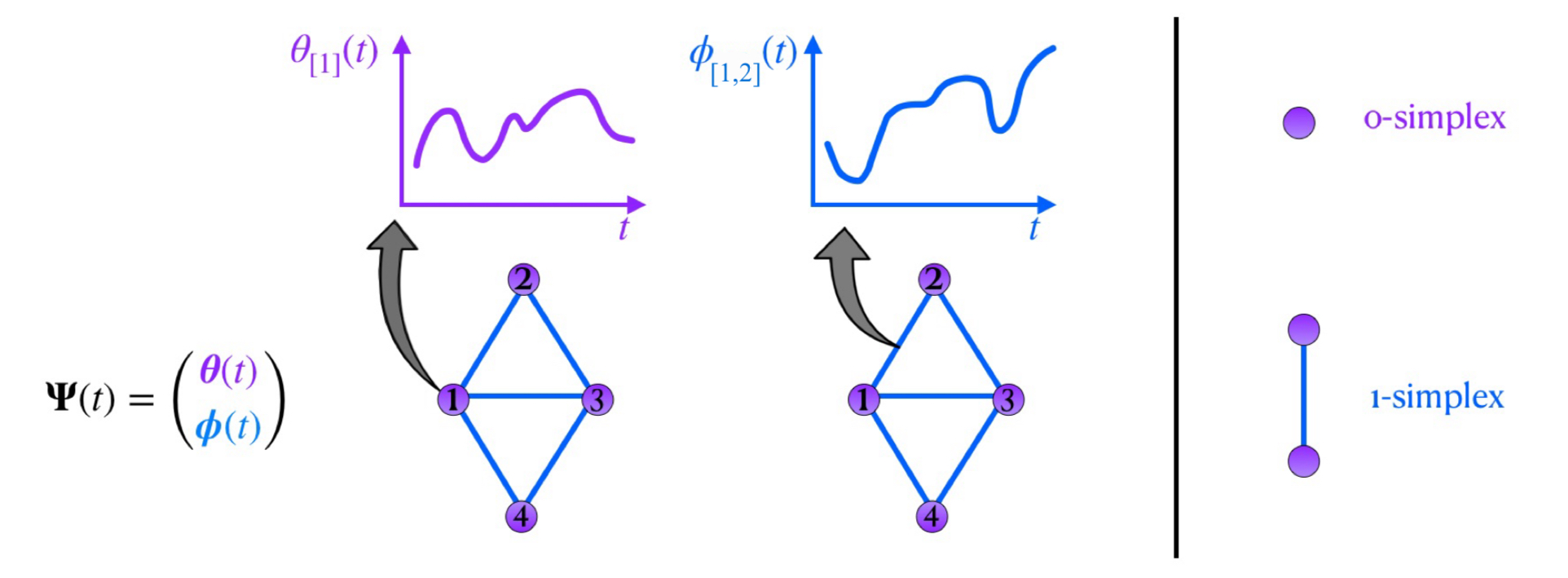}
     \caption{Schematic representation of a network and its dynamical state encoded in the topological spinor $\bm\Psi$ comprising the node signals $\bm\theta\in \mathbb{R}^{N_0}$ and the edge signals $\bm\phi\in \mathbb{R}^{N_1}$. Here the node signal  associated with  node $i$ (0-simplex) of the network is indicated with $\theta_{[i]}(t)$, while the edge signal  associated with each edge $\ell=[i,j]$ (1-simplex) of the network is indicated with $\phi_{\ell}(t)$.}
     \label{fig1}
 \end{figure*}
Here we propose the Dirac-Equation Synchronization Dynamics (DESD) that is formulated starting from a free energy whose fundamental state is designed in order to achieve cluster synchronization aligned with given eigenstates of the Topological Dirac Equation~\cite{bianconi2021topological}.
These eigenstates are defined on both nodes and edges and lead to the topological cluster synchronization state of DESD in which the node and edge frequencies  are determined by the eigenstates of the Topological Dirac Equation.
Therefore, the DESD allows the design of the topological cluster synchronization patterns by modifying the coupling terms among the phase oscillators \cite{stankovski2017coupling} without changing the network topology. This procedure is common in physical approaches to optimization. In these approaches,  the ground state of the Hamiltonian represents the solution of the optimization problem that one desires to solve, and by minimising the Hamiltonian one obtains the optimized solution. For instance, this procedure is common to both simulated annealing \cite{kirkpatrick1983optimization} and quantum annealing \cite{das2008colloquium}.
Similarly, here the coupling term between the phases is derived from the free energy whose fundamental state encodes the topological cluster synchronization patterns that we wish to design. In this way, the proposed approach is able to induce different topological cluster synchronization patterns on the same network by simply changing the parameters of the dynamical model.

Here, using linear stability analysis \cite{millan2019synchronization,millan2018complex,hong2005collective,hong2007entrainment} and adapting it for the study of  topological signals,  we establish the conditions under which the topological cluster synchronization patterns are stable. Moreover,  we demonstrate the ability of DESD to induce several stable topological cluster synchronization patterns on the same network without changing its topology and symmetries. 

We found that DESD leads to stable patterns of the topological cluster synchronization if the eigenstates of the Topological Dirac operator are isolated.  These states can be found, for instance, in random graphs and in stochastic block models (SBM).
In particular, on the SBM we show that we can exploit the important relation between the spectral properties of these networks and their community structure \cite{capocci2005detecting,guimera2004modularity,von2007tutorial,hata2017localization} to induce different patterns of cluster synchronization of nodes and edges that correlate with the modular structure of the network. In general, these states can also be related to network symmetries \cite{sanchez2020exploiting}  through the spectral properties of the networks. As an illustration of the wide applicability of this approach, we studied the DESD on a real connectome derived from diffusion-weighted MRI~\cite{vskoch2022human}, demonstrating that this approach can be used to  reveal the left-right hemispheric separation as well as anterior-posterior differentiation.


Note that this research is based on higher-order topological dynamics \cite{millan2025topology,bianconi2021higher}, a research direction that is distinct from the field that aims to detect edge currents with mechanisms similar to those in condensed matter \cite{tang2021topology,zheng2024topological,chowdhury2025topologically}.However, it is not excluded that these two research directions could be related to each other in the future. 

This paper is structured as follows: In Sec. II we provide a motivation of our work and discuss the relevance of our approach to design topological cluster synchronization in real data, discussing in particular the application of the DESD dynamics to study a brain connectome dataset. In Sec. III we provide the necessary background on the Topological Dirac operator  and on the Kuramoto-like Dirac Topological Synchronization defined on nodes and edges of a network. In Sec. IV we discuss the major properties of the Topological Dirac Equation and its eigenstates. In Sec. V we propose DESD which is able to design the patterns of topological cluster synchronization. We provide a linear stability analysis to assess the stability of these patterns, and provide numerical evidence of topological cluster synchronization on random graphs and on SBM. Finally, in Sec. VI we provide the concluding remarks.
The paper is accompanied by a set of Appendices that provide the details of the linear stability analysis.
\begin{figure*}[!htb!]
     \centering
     \includegraphics[width=1\textwidth]{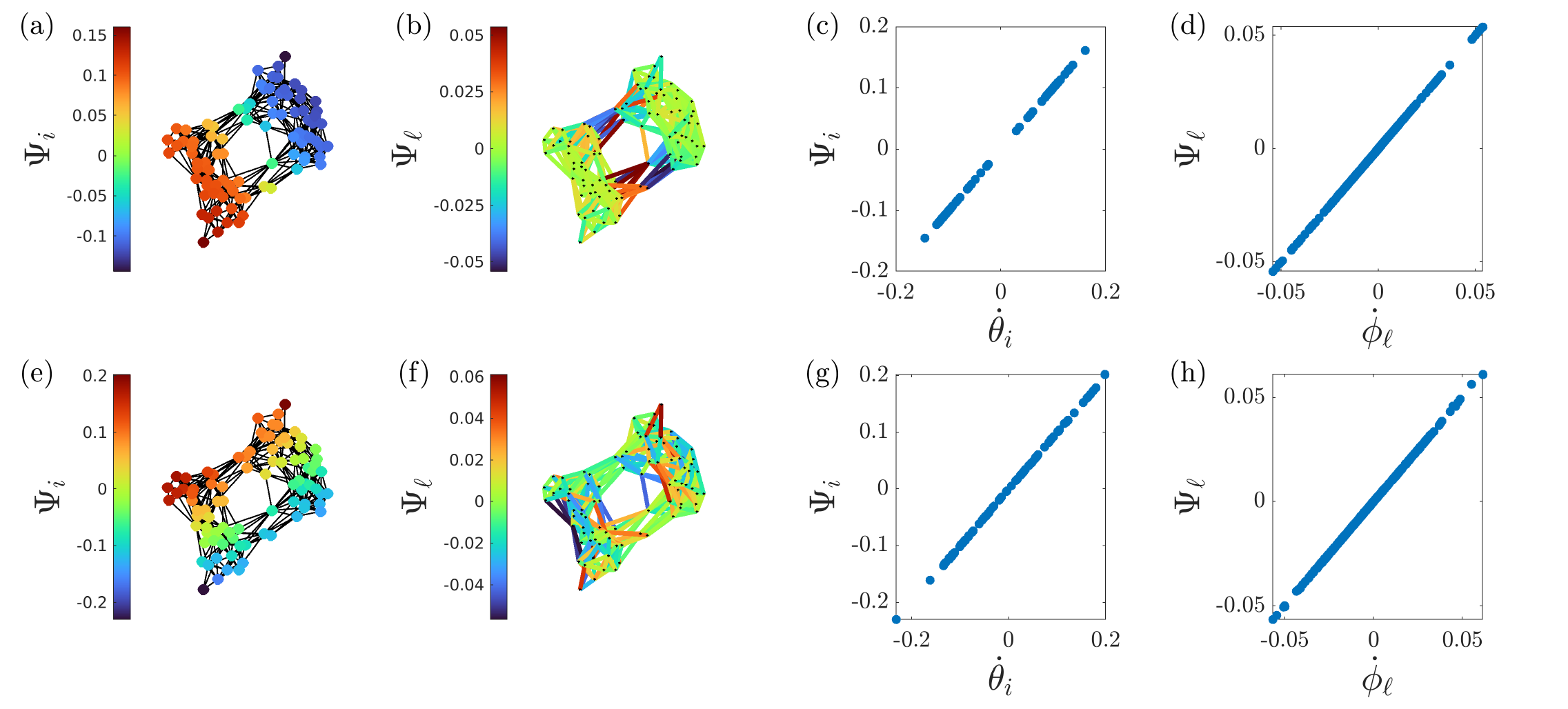}
     \caption{Illustration of the application of DESD to the brain structural connectome  obtained by diffusion-weighted MRI in Ref.\cite{vskoch2022human}. On this network, DESD allows to design two stable topological cluster synchronization patterns aligned along  the first two isolated eigenstate of the Topological Dirac Equation $\bar{E}=1.2392\ldots$ (panels (a)-(d)) and $\bar{E}=1.5053\ldots$ (panels (e)-(h)), of the network. These two topological cluster synchronization patterns partition nodes (panels (a) and (e)) and edges (panels (b) and (f) of the connectome revealing its symmetries and modular structure. In panels (a) and (e), the nodes are colored according to the associated component $\Psi_i^{(\bar{E})}$ of the selected eigenstate of the Topological Dirac Equation revealing in panel (a), the left (red) and right (blue) hemispheres, and in panel (e) the anterior (red) and posterior (blue) parts. In panels (b) and (f), the edges are colored according to the component $\Psi_{\ell}^{(\bar{E})}$ of the selected  eigenstate of the Topological Dirac Equation revealing distinct fiber bundles mediating interactions between these functional modules. Panels (c), (d), (g), and (h) show that the node frequencies $\dot{\theta}_{i}$ and edge frequencies $\dot{\phi}_{\ell}$ associated with the DESD are proportional to the components $\Psi_i^{(\bar{E})}$ and $\Psi_\ell^{(\bar{E})}$ of the selected eigenstate of the Topological Dirac Equation. The simulation of DESD is performed for a mass $m=1$ and a coupling constant $\sigma=15$.}
     \label{fig:brain_network}
\end{figure*}
\section{Overview and of  Dirac-Equation Synchronization Dynamics (DESD)}
\label{preamble}
In this work we embrace a  topological approach to network dynamics\cite{millan2025topology} that associates dynamical variables, also called topological signals,  to both nodes and edges of a network (see Figure \ref{fig1}). Anticipating the mathematical background that will be explained in detail in Sec. \ref{background} we  encode the topological signals of a network in a topological spinor $\bm\Psi=(\bm\theta,\bm \phi)^{\top}$ where 
the node signals and the edge signals are captured by the vectors $\bm\theta\in \mathbb{R}^{N_0}$ and  $\bm\phi\in \mathbb{R}^{N_1}$ respectively, where $N_0$ is the number of the nodes of the network and $N_1$ is the number of the edges. 

Examples of topological signals are ubiquitous as the edge signals can be generally used to treat currents and flows along the edges, as exemplified for instance by the classical example of an electric circuit in which one associates a voltage to each node and a current to each edge of the network. More recently, however, edge signals have been shown to undergo collective phenomena revealing important topological aspects of higher-order network dynamics ~\cite{millan2020explosive,millan2025topology} and they have been investigated  in many  different interdisciplinary contexts, including  brain research~\cite{faskowitz2022edges,santoro2023higher},  and  topological AI algorithms~\cite{barbarossa2020topological,schaub2021signal}.

Note that the topological approach to network dynamics that we will adopt in this manuscript, cannot be directly reduced to dynamics on the line graph of the network~\cite{evans2009line,evans2010line}, where nodes represent the edges of the original network. The reasons for this are twofold. First of all, in the topological approach the dynamics is defined at the same time on both nodes and edges, while the use of the line graph would require to define the dynamics exclusively on the edges of the original network. Secondly, the adoption of the topological approach ensures that the  topology of the networks is fully taken into account and that the approach can be extended {safely}  to higher-order networks (simplicial complexes). In particular, the topological approach preserves the notion of harmonic modes and their connection to the Betti numbers of the network, which would otherwise not be accounted for correctly, as extensively discussed in Ref. \cite{schaub2021signal}.

In this work we will define the Dirac-Equation Synchronization Dynamics (DESD) that will allow us to design topological cluster synchronization on both nodes and edges of a network. This topological cluster synchronization state is designed in such a way to be aligned to isolated eigenstates of the Topological Dirac Equation \cite{bianconi2021topological} which are defined on both nodes and edges of the network and reveal its modular structure and symmetries.
Before we discuss in detail the theoretical framework that will allow us to draw different topological cluster synchronization states without modifying the network structure, let us motivate our research question by illustrating a direct application of our theoretical framework to a real world scenario.

Here we demonstrate the applicability of the DESD to brain research, by investigating this dynamical process  on  the structural connectivity of a human brain network derived from diffusion-weighted MRI \cite{vskoch2022human} (Figure \ref{fig:brain_network}). The isolated eigenstates of the Topological Dirac Equation, represented by the color code of node and edges in the network visualizations shown in panels (a)-(b) and (e)-(f), naturally reveal macroscale partitions that correspond to well-established functional subdivisions: most prominently, a left–right hemispheric separation and an anterior–posterior differentiation. When the DESD is designed in such a way to recover a topological cluster synchronization state aligned along these eigenstates,  the frequencies of the node and of the edge signals, given by $\dot{\bm\theta}$ and $\dot{\bm\phi}$ respectively, induce topological patterns of cluster synchronization aligned with these anatomical modules, illustrating how structural topology constrains coherent functional activity. This is evident from  the clear proportionality between the frequencies of nodes $\dot{\theta}_i$ and edges $\dot{\phi}_\ell$ and, respectively, the component of the eigenstate of the Topological Dirac Equation defined on nodes $\Psi^{(\bar{E})}_i$ and edges $\Psi^{(\bar{E})}_\ell$ of the network obtained for sufficiently large coupling constant of the DESD (panels (c)-(d) and (g)-(h)). 
 Thus, when the DESD designs the topological cluster synchronization along the respective eigenstate of the Topological Dirac Equation, the dynamics on both nodes and edges reveals the left and right hemispheres and the segmentation between the anterior and posterior regions of the brain. These results  show that the DESD provides a natural bridge between anatomical connectivity and emergent functional organization, with cluster synchronization patterns arising coherently in both nodes and edges—something not captured by classical node-based dynamical models. Specifically,  DESD identifies physically interpretable synchronization modes linking structural and functional brain modularity, thus addressing the key challenge of bridging topology and dynamics in real neural systems.


We note that DESD allows to design multiple topological cluster synchronization states without changing the topology or the symmetries of the network. Interestingly, this phenomenology is also achieved without changing the geometry of the network which is naturally captured by the edge weights. Recently, weights have been  identified as a powerful tool to modify the dynamical properties of a network or a simplicial complex  in the context of node-based cluster synchronization \cite{duan2022prevalence}, global (topological) synchronization \cite{wang2024global,gallo2025global}, and diffusion~\cite{baccini2022weighted} and random walks \cite{tian2025matrix}. Here, we embrace a purely topological approach and we adopt trivial edge weights equal to one on every edge. The rich phenomenology that we observe can be further extended including in the definition of the DESD non-trivial edge weights according to the algebraic topology approach to network dynamics \cite{baccini2022weighted}, thereby exploring  the effect of network geometry in DESD. However, this extension is beyond the scope of this work and will be addressed in future investigations.

Having now provided an overview of DESD and having demonstrated its applicability to real brain networks, in the rest of the manuscript we will adopt a  comprehensive approach to DESD providing the algebraic topology and dynamical system background to this theoretical framework and investigating the stability of its associated topological cluster synchronization dynamics.

\section{Background on Dirac Topological Synchronization}
\label{background}
\subsection{The Topological Dirac operator of a network}

We consider a network $G=(E,V)$ consisting of a set of $N_{0}$ nodes $V=\{1,2,\dots, i, \ldots, N_0\}$ and a set $E$ of $N_{1}$ edges $E=\{1,2,\ldots, \ell,\ldots, N_1\}$. Here and in the following, we will indicate with $\mathcal{N}$ the total number of simplices in the network (nodes and edges) and with $r\in \{1,2,\ldots, \mathcal{N}\}$ the generic simplex. 
In the field of higher-order topological dynamics \cite{millan2025topology,bianconi2021higher} the dynamical state of the network (see Figure $\ref{fig1}$) is encoded by the {\it topological spinor} $\bm\Psi\in \mathbb{R}^{\mathcal{N}}$ given by the direct sum of the node signal $\bm\theta\in \mathbb{R}^{N_0}$ and the edge signal $\bm\phi\in\mathbb{R}^{N_1}$, i.e.
\bea
\bm\Psi=\begin{pmatrix}\bm\theta\\
\bm\phi\end{pmatrix}.\label{spinor}
\eea
 The Topological Dirac operator ${\bf D}$ \cite{bianconi2021topological}  is a fundamental algebraic topology operator for performing discrete exterior calculus on the topological spinor. It is defined as 
\begin{equation*}
    {\bf D}=
    \begin{pmatrix}
        {\bf 0}&{\bf B}\\
        {\bf B}^{\top}&{\bf 0}\\
    \end{pmatrix}
\end{equation*}
where the incidence matrix ${\bf B}$  is an $N_{0}\times N_{1}$ matrix of elements
\begin{equation*}
    {B}_{i\ell}=
    \begin{cases}
        1 & \text{if $\ell=[j,i]$, $j<i$}\\
        -1 & \text{if $\ell=[i,j]$, $i<j$}\\
        0 & \text{otherwise}.
    \end{cases}
\end{equation*}
Note that the incidence matrix ${\bf B}$ is a matrix representation of the boundary operator, mapping each edge to its end nodes.
The Dirac acts on the topological spinor as
\bea
{\bf D}\bm\Psi=\begin{pmatrix}{\bf B}\bm\phi\\{\bf B}^{\top}\bm\theta\end{pmatrix}.
\eea
Thus, it projects the node signals onto the edge signals and the edge signals onto the node signals thereby allowing cross-talk between them.
Let us now discuss in more detail the nature of this transformation, which is deeply rooted in discrete exterior calculus \cite{millan2025topology,bianconi2021higher,grady2010discrete}.
 The projection of the node signals onto edge signals ${\bf B}^{\top}\bm\theta\in \mathbb{R}^{N_1}$ is nothing other than the gradient of the node signals and has elements
\bea
[{\bf B}^{\top}\bm\theta]_{\ell}=\theta_j-\theta_i,
\eea  
given by the difference in the value of $\theta$ at the two end nodes of the edge $\ell=[j,i]$.
Moreover, the projection of the edge signals onto the node signals  ${\bf B}\bm\phi\in \mathbb{R}^{N_0}$ is given by  the divergence of the edge signals and has elements
\bea
[{\bf B}\bm\phi]_{i}=\sum_{\ell=[j,i]}\phi_{\ell}-\sum_{\ell=[i,j]}\phi_{\ell}
\eea
expressing the  difference between the flux in and the flux out of node $i$.

A key property of the Dirac operator \cite{bianconi2021topological} is that it can be treated as the `square root' of the Laplacian. Indeed, we have ${\bf D}^{2}=\bm{\mathcal{L}}$ where $\bm{\mathcal{L}}$ is the Gauss-Bonnet Laplacian given by:
\begin{equation}
    \bm{\mathcal{L}} =
    \begin{pmatrix}
        {\bf L}_{[0]} & {\bf 0}\\
        {\bf 0} & {\bf L}_{[1]}
    \end{pmatrix}
\end{equation}
where \bea
{\bf L}_{[0]}={\bf B}{\bf B}^{\top}\quad {\bf L}_{[1]}={\bf B}^{\top}{\bf B}.
\label{L01}
\eea
Here ${\bf L}_{[0]}$ is the famous graph Laplacian of the network, describing diffusion from node to node passing through edges, while ${\bf L}_{[1]}$ is the 1-Hodge Laplacian describing the diffusion from edge to edge passing through nodes. Note that here and in the following we indicate with $\lambda>0$ the singular value of ${\bf B}$ and with ${\bf u}_{\lambda},{\bf v}_{\lambda}$ its corresponding left and right singular vectors. Given the definition (Eq.(\ref{L01})) of the graph Laplacian ${\bf L}_{[0]}$ and the 1-Hodge Laplacian ${\bf L}_{[1]}$, it follows that ${\bf L}_{[0]}$ and ${\bf L}_{[1]}$ are isospectral, i.e., they have the same non-zero eigenvalues $\mu=\lambda^2$, but generally have different degeneracies for the harmonic eigenvalues, given by the Betti numbers $\beta_0$ and $\beta_1$, respectively. The $n^{th}$ Betti number is a topological invariant equal to the number of $n$-dimensional holes in the network. In particular, $\beta_{0}$ is equal to the number of connected components and $\beta_{1}$ is equal to 1-dimensional holes (i.e. cycles). In a connected network, we thus have $\beta_0=1$ and $\beta_1=N_1-(N_0-1)$.

Given that ${\bf D}^2=\bm{\mathcal{L}}$, it can be shown \cite{bianconi2021topological,calmon2023dirac} that the eigenvalues $\Lambda$ of ${\bf D}$ are given by 
\bea
\Lambda=\pm\lambda=\pm \sqrt{\mu}
\eea
where $\mu$ is the generic eigenvalue of ${\bf L}_{[0]}$. 
Additionally, the eigenvectors of ${\bf D}$ can be encoded in the matrix:
\bea
    \left(\begin{array}{cccc}{\bf W}^{-} &  {\bf W}_{\textrm{harm}}^{-} &{\bf W}_{\textrm{harm}}^{+}&{\bf W}^{+}\end{array}\right).   
\eea
Here, ${\bf W}^{\pm}$ are the matrices associated with the eigenvectors $\bm\Psi^{(\Lambda)}$ with eigenvalues $\Lambda>0$ and eigenvalues $\Lambda<0$, respectively. These eigenvectors can be explicitly expressed as  
\bea
    \boldsymbol{\Psi}^{(\Lambda)}&=&\frac{1}{\sqrt{2}}\left(\begin{array}{c}
        {\bf u}_\lambda \\
         {\bf v}_\lambda
    \end{array}\right)\quad\ \mbox{for}\  \Lambda>0,\nonumber \\
    \boldsymbol{\Psi}^{(\Lambda)}&=&\frac{1}{\sqrt{2}} \left(\begin{array}{c}
     {\bf u}_\lambda \\
       -{\bf v}_\lambda
    \end{array}\right)\quad\mbox{for}\  \Lambda<0,
    \eea
    where ${\bf u}_\lambda$ and ${\bf v}_{\lambda}$ are the left and right singular vectors of the boundary operator ${\bf B}$ associated with the singular value $\lambda$, and $\mathcal{C}^{\pm}$ are normalization constants.

    The matrices $\bf W_{\textrm{harm}}^{\pm}$ encode the harmonic eigenvectors associated with $\Lambda=0$. These eigenvectors are $\beta_0+\beta_1$ degenerate and there is a basis in which they have the following structure
\bea
    \boldsymbol{\Psi}^{(0)}_N&=&\left(\begin{array}{c}
        {\bf u}_0 \\
        {\bf 0}
    \end{array}\right), \quad\mbox{with degeneracy}\  \beta_0,
\nonumber \\    \boldsymbol{\Psi}^{(0)}_L&=&\left(\begin{array}{c}
        {\bf 0} \\
        {\bf v}_0 \\
    \end{array}\right),  \quad\mbox{with degeneracy}\  \beta_1.
    \label{Psi0}
\eea
As previously mentioned, in a connected network $\beta_0=1$ and its corresponding eigenvector ${\bf u}_0\propto{\bf 1}_{N_0}$ is constant over each node of the network, while ${\bf v}_0$ is the generic harmonic eigenvector on the edges (harmonic eigenvectors of ${\bf L}_{[1]}$). The eigenvectors ${\bf v}_0$ are as many as the independent cycles (i.e., $\beta_1=N_1-(N_0-1)$ degenerate), and there is a basis in which each of them localizes around a single independent cycle of the network.


\subsection{Dirac Topological Synchronization (DTS)}
Topological synchronization is a collective synchronization phenomenon that involves topological signals. Thus, on a simple network it involves both node and edge signals. Here we consider Topological synchronization of Kuramoto-like \cite{kuramoto1975self,pikovsky2001synchronization,rodrigues2016kuramoto,arenas2008synchronization,boccaletti2018synchronization} models involving non-identical oscillators of phases $\bm\theta$ and $\bm\phi$ associated with the nodes and edges of the network, respectively. Thus, the dynamical state of the network is encoded in the topological spinor defined in Eq. (\ref{spinor}), and the node and edge topological oscillators are coupled via the Dirac operator.
In the absence of interactions ($\sigma=0$), the dynamical equations describe non-identical oscillators associated with nodes and edges, each oscillating with a heterogeneous intrinsic frequency. Specifically, we have
\bea 
    \frac{d{{\bm\Psi}}}{dt}={\bm\Omega}=\begin{pmatrix}\bm\omega\\\bm{\hat{\omega}}\end{pmatrix},
\eea
where $\bm\omega\in \mathbb{R}^{N_0}$  and $\bm{\hat{\omega}}\in\mathbb{R}^{N_1}$ are the vectors of intrinsic frequencies associated with the nodes and edges, respectively.
When interactions are turned on ($\sigma>0$), the equations for the Dirac Topological Synchronization (DTS) are
\bea 
    \frac{d{{\bm\Psi}}}{dt}={\bm\Omega}-\frac{\delta \mathcal{F}}{\delta {{\bm\Psi}}}
    \label{PsiF}
\eea
where $\mathcal{F}$, here called the {\em free energy}, encodes for the interactions among phases associated with nodes and edges. In its simplest form $\mathcal{F}$ is given by \cite{millan2025topology,arnaudon2022connecting}
\bea
\mathcal{F}=-\sigma {\bf 1}^{\top}\cos({\bf D}\bm\Psi)
\label{F}
\eea
where the cosine function is taken element-wise and ${\bf 1}$ is the $\mathcal{N}$-column vector whose elements are $1_r=1$.
With this choice of the free energy $\mathcal{F}$, we obtain the DTS equation of motion given by
\bea \label{DTS}
    \frac{d{{\bm\Psi}}}{dt}={\bm\Omega}-\sigma{\bf D}\sin({\bf D}\bm\Psi)
\eea
where the sine function is taken element-wise. This system of equations can also be expressed as
\bea
    \frac{d{\bm\theta}}{dt}&=&{\bm\omega}-\sigma{\bf B}\sin({\bf B}^{\top}{\bm\theta}),\label{K1}\\
    \frac{d{\bm \phi}}{dt}&=&\hat{{\bm\omega}}-\sigma{\bf B}^{\top}\sin({\bf B}{\bm\phi}).\label{K2}\eea
 The first thing we notice from these equations is that, with this choice of free energy, the node signal $\bm\theta$ and the edge signal $\bm\phi$ are decoupled. Equation (\ref{K1}) for the node signals $\bm\theta$ can be shown (see Ref.\cite{millan2020explosive}) to be the standard Kuramoto model defined on the network $G$, while Eq. (\ref{K2}) for the edge signals $\bm\phi$ is the topological synchronization proposed and studied in Refs. \cite{millan2020explosive,ghorbanchian2021higher}.

The fully synchronized state of DTS corresponds to the ground state of the free energy $\mathcal{F}$ given by Eq.(\ref{F}) and satisfies 
\bea
{\bf D}\bm\Psi=0
\eea
This implies that $\bm\Psi\in \mbox{ker}({\bf D})=\mbox{ker}({\boldsymbol{\mathcal{L}}})$ and thus, in a fully connected network, implies that $\bm\Psi$ is aligned along a linear combination of $\bm\Psi^{(0)}_N$ and $\bm\Psi^{(0)}_L$, which are  given by Eq.(\ref{Psi0}).
Assuming that neither $\bm\theta$ nor $\bm\phi$ are zeros, this implies that $\bm\theta$ is the same on each node, i.e., $\bm\theta={\bf 1}$ and $\bm\phi$ is aligned along the harmonic eigenvectors of the network, i.e., it is localized along the cycles of the network.

 Interestingly, the linearized dynamics of DTS Eq.(\ref{DTS}), valid when  ${\bf D}\bm\Psi$ is small, is given by 
 \bea
  \frac{d{{\bm\Psi}}}{dt}={\bm\Omega}-\sigma{\boldsymbol{\mathcal{L}}}\bm\Psi,
 \eea
 which can be also written explicitly as 
 \bea
 \frac{d{\bm\theta}}{dt}&=&{\bm\omega}-\sigma{\bf L}_{[0]}\bm\theta,\label{KL1}\\
   \frac{d{\bm \phi}}{dt}&=&\hat{{\bm\omega}}-\sigma{\bf L}_{[1]}{\bm\phi}.\label{KL2} \eea
 The linearized Kuramoto model Eq.(\ref{KL1}) has been studied on networks \cite{millan2019synchronization,millan2018complex} and lattices \cite{hong2005collective,hong2007entrainment} to investigate the stability of the fully synchronized state, revealing the role of a finite spectral dimension in determining whether the fully synchronized state of the nodes can be observed. Note, however, that the linear stability of the topological synchronization state of edge signals starting from Eq.(\ref{KL2}) in the presence of a finite spectral dimension has not been investigated so far.

 Going beyond the linear approximation regime, and embracing the full nonlinear nature of the DTS, we can investigate the behavior of the dynamics by separately studying the harmonic and non-harmonic component of $\bm\Psi$. 
 To this end, we observe that $\bm\Psi$ can be expressed in a unique way as 
\bea
    \bm\Psi=\bm\Psi^{\text{harm}}+\bm\Psi^{(D)}
\eea
where $\bm\Psi^{\text{harm}}\in \mbox{ker}({\bf D})$ and $\bm\Psi^{(D)}\in \mbox{im}({\bf D})$ and that similarly the vector of intrinsic frequencies $\bm\Omega$ can be decomposed into 
\bea
    \bm\Omega=\bm\Omega^{\text{harm}}+\bm\Omega^{(D)}
\eea
 where $\bm\Omega^{\text{harm}}\in \mbox{ker}({\bf D})$ and $\bm\Omega^{(D)}\in \mbox{im}({\bf D})$.
By using this decomposition, it can be readily shown that the harmonic component  $\bm\Psi^{\text{harm}}$  in the full nonlinear dynamics of DTS is decoupled and non-interacting, and obeys
\bea 
    \frac{d{{\bm\Psi}}^{\text{harm}}}{dt}={\bm\Omega}^{\text{harm}}
    \label{free}
\eea
Thus this harmonic component continues to oscillate unaffected by the nonlinear dynamics. 
In order to investigate the dynamics of the non-harmonic component of the signal, i.e. $\bm\Psi^{(D)}$ we consider 
\bea
    {\bm\Theta}={\bf D}\bm\Psi={\bf D}\bm\Psi^{(D)}.
    \label{Theta}
\eea
By expressing the Topological Dirac operator ${\bf D}$ explicitly, we can express the node and edge components of $\bm\Theta$ as 
\bea
\bm\Theta=\begin{pmatrix}
        {\bm\alpha}\\
        {\bm\beta}
    \end{pmatrix}=\begin{pmatrix}
        {\bf 0}&{\bf B}\\
        {\bf B}^{\top}& {\bf 0}
    \end{pmatrix}\begin{pmatrix}
        {\bm\theta}\\
        {\bm\phi}
    \end{pmatrix}.
\eea
Thus, $\bm\Theta$ is unaffected by the harmonic component of $\bm\Psi$ and comprises the projection of the edge signal onto the nodes, $\bm\alpha$, and the projection of the node signal onto the edges, $\bm\beta$.
Using Eq.(\ref{DTS}) and the definition of $\bm\Theta$ (Eq.(\ref{Theta})), we can easily obtain  the dynamical equation of motion for $\bm\Theta$ that are given by 
\bea
    \frac{d{{\bm\Theta}}}{dt}={\bf D}{\bm\Omega}-\sigma{\boldsymbol{\mathcal{L}}}\sin({\bm\Theta}),
\eea
or equivalently by
\bea
    \frac{d{{\bm\alpha}}}{dt}&=&{\bf B}\hat{\bm\omega}-\sigma{\bf L}_{[0]}\sin({\bm\alpha}),\nonumber \\
    \frac{d{\bm \beta}}{dt}&=&{\bf B}^{\top}{\bm\omega}-\sigma{\bf L}_{[1]}\sin({\bm\beta}).
\eea
From the study of the dynamics of $\bm\Psi^{\text{harm}}$ and $\bm\Theta={\bf D}\bm\Psi^{(D)}$ the following physical understanding of the DTS dynamics emerges.
The fully synchronized state of DTS is the state in which $\bm\Theta$ freezes and for which only the dynamics of $\bm\Psi^{\text{harm}}$ remains described by Eq.(\ref{free}). This occurs in general for high values of $\sigma$.
Under these conditions, the node phases oscillate in unison as their corresponding harmonic eigenvector is ${\bf u}_0\propto{\bf 1}$, while the edge phases oscillate by remaining aligned with the harmonic eigenvectors ${\bf v}_0$ and thus have dynamics that localize around 1-dimensional holes (i.e., cycles). Thus, the localization of edge signals around holes can be interpreted as the dynamics learning an aspect of the network topology.

For evaluating how well the synchronized dynamics of the projected phases freeze, two order parameters have been proposed, $R_{\alpha}$ and $R_{\beta}$, given by
\bea
    R_{\alpha} = \left|\frac{1}{N_{0}}\sum_{i=1}^{N_0}e^{i\alpha_{i}}\right|, \quad
    R_{\beta} = \left|\frac{1}{N_{1}}\sum_{\ell=1}^{N_1}e^{i\beta_{\ell}}\right|.
\eea
These order parameters display on random graphs a smooth behavior, with a continuous onset of the non-zero values occurring at $\sigma^{\star}=0$. 

The study of the DTS opens very interesting perspectives on the dynamics of topological oscillators. In particular, here,  we have shown that DTS leaves node and edge signals decoupled. Thus, an important problem is to define dynamics that couples oscillators placed on nodes and edges.
In the framework of Kuramoto-like models, this has been the subject of growing activity, and includes approaches leading to discontinuous phase transitions demonstrated by the behavior of $R_{\alpha}$ and $R_{\beta}$. These models directly modify the dynamical Eq. (\ref{DTS}) for DTS and cannot be written in terms of Eq.(\ref{PsiF}), i.e., they are not associated with a free energy $\mathcal{F}$. The first approach proposes to modulate the coupling constant $\sigma$ with the order parameters \cite{millan2020explosive,ghorbanchian2021higher}. The second approach, also called Topological Dirac synchronization \cite{calmon2022dirac,calmon2023local,nurisso2024unified}, instead introduces a phase lag into the equation of motion which then reads 
\begin{equation} \label{dirac_v0}
    \frac{d{{\bm\Psi}}}{dt}={\bm\Omega}-\sigma\hat{\bf D}\sin((\hat{\bf D}-z{\bm\gamma}\hat{\bf D}^2)\bm\Psi)
\end{equation}
where $z\in (0,1)$, where $\hat{\bf D}$ is a suitably defined normalized Dirac operator and the matrix $\bm\gamma$ is given by
\bea
    {\bm\gamma}=
    \begin{pmatrix}
        {\bf I}_{N_0}&{\bf 0}\\
        {\bf 0}&-{\bf I}_{N_1}\\
    \end{pmatrix}
    \label{gamma0}
\eea

Until now, we have shown that in DTS the synchronization dynamics occurs along the harmonic eigenvectors of the Topological Dirac operator ${\bf D}$ that are constant on the nodes and localize on edges around cycles. Thus, an important question emerges: can we design a dynamical system that synchronizes along other network patterns? 

To this end, we will introduce the Dirac-Equation Synchronization Dynamics (DESD) which will affect the coupled dynamics of node and edge phases and will allow us to design spatial patterns on the network along which we can localize the synchronized dynamics. This synchronization model will exploit the properties of the Topological Dirac Equation. Hence, before discussing DESD, we first introduce the basic properties of the Topological Dirac Equation.

\section{The Topological Dirac Equation}
The Topological Dirac Equation \cite{bianconi2021topological} is a dynamical equation for the topological spinor that gives rise to a relativistic dispersion relation of its energy states, much like the Dirac equations defined on the continuum. This equation extends the traditional approach for staggered fermions on the lattice \cite{kogut1975hamiltonian} to any arbitrary network structure and has been recently shown to have wide applications in AI algorithms, including signal processing \cite{calmon2023dirac,wang2025dirac} and graph neural network algorithms \cite{nauck2024dirac}.
The Topological Dirac Equation is defined as
\bea \label{diraceqn}
    \textrm{i}\frac{d\bm\Psi}{dt}=\boldsymbol{\mathcal{H}}\bm\Psi 
    \eea
    where the Hamiltonian $\boldsymbol{\mathcal{H}}$ is given by 
    \bea
    \boldsymbol{\mathcal{H}}={\bf D}+m\bm\gamma.
\eea
Here $m\geq0$ is a parameter called {\it the mass} and the gamma matrix ${\bm\gamma}$ is given by Eq.(\ref{gamma0}).

\begin{figure}
     \includegraphics[width=0.97\columnwidth]{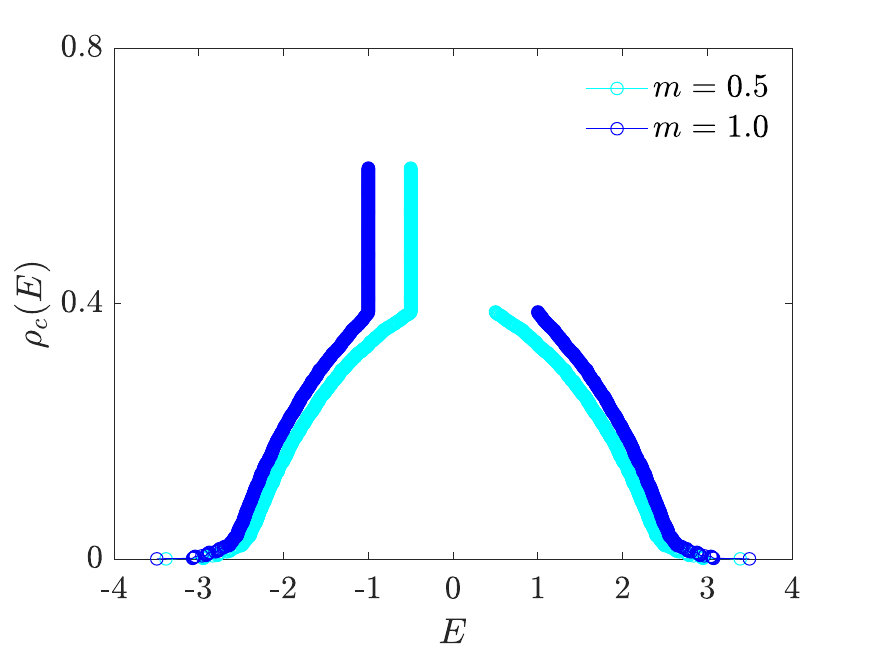}
     \caption{The spectrum of the Topological Dirac Equations is described by  the cumulative distribution of eigenstates $\rho_c(E)$ indicating for positive energy values $E$ the number of eigenstates of energy $E^{\prime}$ with $E^{\prime}\geq E$ and for negative  energy values  $E$ the number of eigenstates of energy $E^{\prime}$ with $E^{\prime}\leq E$. From this plot, two main observations can be made: (i) the non-zero mass $m$ induces a spectral gap; (ii) the spectrum is symmetric with the only exception of the eigenstates of energy $E$ with $|E|=m$. The data shown here is for a random network with $N_{0}=492$ nodes and average degree $c=3$.}
     \label{spectra}
 \end{figure}

Note that according to this definition, the gamma matrix $\bm\gamma$ anti-commutes with the Dirac operator, i.e., $\{{\bf D},{\bm\gamma}\}={\bf D}{\bm\gamma}+{\bm\gamma}{\bf D}=0$.
The solution of the Topological Dirac Equations is
\bea
    \bm\Psi=e^{-\textrm{i}Et}\bm \Psi^{(E)}
\eea
where $\bm \Psi^{(E)}$ is the eigenstate of the Topological Dirac Equations associated with the {\it energy} $E$. Any eigenstate $\bm \Psi^{(E)}$ is independent of time and satisfies
\bea
    E\bm\Psi^{(E)}=\boldsymbol{\mathcal{H}}\bm\Psi^{(E)}
\eea
or equivalently,  
\bea \label{eq:eig}
    (\boldsymbol{\mathcal{H}}-E{\bf I})\bm\Psi^{(E)}=0.
\eea
Iterating this eigenstate equation twice and using $\{{\bf D},\bm\gamma\}=0$ we find that 
\bea
    E^2\bm\Psi^{(E)}={\bf D}^2\bm\Psi^{(E)}+m^2\bm\Psi^{(E)}
\eea
This relation reveals that the eigenstates of the Dirac operators are eigenstates of the Gauss-Bonnet Laplacian $\mathcal{L}={\bf D}^2$ associated with the eigenvalue $\mu=\lambda^2$ and that the energy $E$ of the Topological Dirac Equations obeys the relativistic dispersion relation 
\bea
    E^2=\lambda^2+m^2.
\eea
From this relation we conclude that the role of the mass is to introduce a gap in the energy spectrum, as the energy values must have an absolute value greater than or equal to the mass, i.e., $|E|\geq m$. A typical energy spectrum of the Topological Dirac Equation is plotted in Figure \ref{spectra}, from which it is apparent that the spectrum includes both positive and negative energy values and every positive energy state with $E>m$ admits a conjugated negative energy state with the same absolute value $|E|$. This symmetry corresponds to the charge conjugation symmetry of the continuous Dirac equation. However, in the discrete case of the Topological Dirac Equation, this symmetry is broken for the energy states of energy $|E|=m$ which correspond to the harmonic eigenstates of the network and thus have a degeneracy given by the Betti number $\beta_0$ (number of connected components) for $E=m$ and a degeneracy given by the Betti number $\beta_1$ (number of independent cycles) for $E=-m$.

The mass not only changes the energy spectrum of the Topological Dirac Equation, but also significantly changes the properties of the eigenstates associated with the energy $E$. 
In order to see this, let us discuss the structure of the eigenstates associated with the energy $E$ and the singular value $\lambda$ of the incidence matrix.  To this end, let us characterize the matrix of the eigenstates 
\bea
    \left(\begin{array}{cccc}\hat{\bf W}^{-} &  \hat{\bf W}_{\textrm{harm}}^{-} &\hat{\bf W}_{\textrm{harm}}^{+}&\hat{\bf W}^{+}\end{array}\right).   
\eea
Here $\hat{\bf W}^{\pm}$ are the matrices associated with  the eigenstate $\bm\Psi^{(E)}$ with  $E>m$ or $E<-m$ respectively (and $\lambda\neq 0$), which are given by 
\bea
    \boldsymbol{\Psi}^{(E)}&=&\mathcal{C}^+\left(\begin{array}{c}
        {\bf u}_\lambda \\
        \frac{ \lambda}{|E|+m} {\bf v}_\lambda
    \end{array}\right),\quad\mbox{for}\  E>m\nonumber \\
    \boldsymbol{\Psi}^{(E)}&=&\mathcal{C}^- \left(\begin{array}{c}
    \frac{ \lambda}{|E|+m} {\bf u}_\lambda \\
       -{\bf v}_\lambda
    \end{array}\right),\quad\mbox{for}\  E<-m
    \label{eq:Dirac_eigestates}
\eea
    where ${\bf u}_\lambda$ and ${\bf v}_{\lambda}$ are the left and right singular vectors of the boundary operator ${\bf B}$ associated with the singular value $\lambda$ and $\mathcal{C}^{\pm}$ are normalization constants. Note that ${\bf u}_\lambda$ and ${\bf v}_{\lambda}$ are also eigenvectors of the graph Laplacian and of the $1$-Hodge Laplacian respectively both associated to the eigenvalue $\lambda^2$. Thus, these vectors reflect the symmetries of the network \cite{sanchez2020exploiting} and its community structure \cite{capocci2005detecting,von2007tutorial}.
    We note that the mass now enables the tuning of the relative normalization of the node and the edge signals, which can now have very different scales. Only for $m=0$ do these eigenvectors reduce to the eigenvectors of the Dirac operator. The matrices $\hat{\bf W}_{\textrm{harm}}^{\pm}$ encode the harmonic eigenstates associated with $\lambda=0$ and energies $E=\pm m$. These eigenvectors are independent of the value of the mass and are given by  
\bea
    \boldsymbol{\Psi}^{(E)}&=&\left(\begin{array}{c}
        {\bf u}_0 \\
        {\bf 0}
    \end{array}\right), \quad\mbox{for}\  E=m
    \nonumber \\    \boldsymbol{\Psi}^{(E)}&=&\left(\begin{array}{c}
        {\bf 0} \\
        {\bf v}_0 \\
    \end{array}\right),  \quad\mbox{for}\  E=-m
    \nonumber
\eea
Note that the degeneracy of the eigenvalue $E=m$ is equal to the degeneracy of the harmonic eigenvalue of ${\bf L}_{[0]}$ and is thus  given by the zeroth Betti number $\beta_0$, while the degeneracy of the eigenvalue $E=-m$ is equal to the degeneracy of the harmonic eigenvalue of ${\bf L}_{[1]}$ and is thus given by the first Betti number $\beta_1$.

 
\section{Dirac-Equation Synchronization Dynamics and its stability }
\subsection{Dirac-Equation Synchronization Dynamics (DESD)}
The Dirac-Equation Synchronization Dynamics (DESD) allows us to design the synchronized state by aligning it to an eigenstate associated with the energy $\bar{E}$ of the Topological Dirac Equation. 
To this end, we consider the dynamics dictated by the equation of motion
\bea 
    \frac{d{{\bm\Psi}}}{dt}={\bm\Omega}-\frac{\delta \mathcal{F}}{\delta {{\bm\Psi}}},
    \label{d2}
\eea
where  the free energy $\mathcal{F}$ is now chosen to be
\bea
    \mathcal{F}=-\sigma {\bf 1}^{\top}\cos((\boldsymbol{\mathcal{H}}-\bar{E}{\bf I})\bm\Psi).
    \label{FD}
\eea
Notice that our choice of the free energy corresponds to the wish to design the synchronization pattern by imposing that the ground state of this free energy is given by the eigenstate associated with the energy $\bar{E}$ of the Topological Dirac Equation. 
Indeed, the free energy $\mathcal{F}$ is minimized for 
\bea
    (\boldsymbol{\mathcal{H}}-\bar{E}{\bf I}){\bm \Psi}={\bf 0},
\eea
    where here and in the following ${\bf I}$ is the identity matrix of linear size $\mathcal{N}$.
Thus, the ground state of the free energy has topological spinors $\bm\Psi$ satisfying the eigenstate equation 
\bea
    \boldsymbol{\mathcal{H}}{\bm \Psi}=\bar{E}{\bm \Psi}.
\eea
It follows that such a ground state only exists if $\bar{E}$ is an allowed energy state of the Topological Dirac Equation.

It is instructive at this point to provide the explicit expression for $\boldsymbol{\mathcal{H}}-\bar{E}{\bf I}$
\bea
    (\boldsymbol{\mathcal{H}}-\bar{E}{\bf I})=\begin{pmatrix}
        -(\bar{E}-m){\bf I}_{N_0}&{\bf B}\\
        {\bf B}^{\top}& -(\bar{E}+m){\bf I}_{N_1}
    \end{pmatrix}
\eea
revealing that this operator allows non-trivial cross-talk between nodes and edge signals.
Inserting  Eq.(\ref{FD}) into Eq.(\ref{d2}) we can thus derive the dynamical equations of motion of DESD which read 
\bea
    \frac{d\bm\Psi}{dt}=\bm\Omega-\sigma(\boldsymbol{\mathcal{H}}-\bar{E}{\bf I})\sin((\boldsymbol{\mathcal{H}}-\bar{E}{\bf I})\bm\Psi),
    \label{DESD}
\eea
Therefore, the DESD in general describes coupled dynamics of node and edge signals, although it still obeys Eq.(\ref{d2}).
Note, however, that for $m=\bar{E}=0$ we recover the  Dirac Topological Synchronization (DTS) as Eq.(\ref{DESD}) reduces to 
\bea
    \dot{\bm\Psi}=\bm\Omega-\sigma{\bf D}\sin({\bf D}\bm\Psi).
\eea
for which node and edge signals are decoupled.
Here and in the following, we consider exclusively the case in which $\bar{E}$ is a non-degenerate energy state of the Topological Dirac Equation. Moreover the intrinsic frequencies $\bm\Omega$ are given by 
\bea
    \bm\Omega=\sum_{E}\Omega_{E}(t)\bm\Psi^{(E)}
\eea
with $\Omega_E$ drawn from a standard Gaussian distribution for $E\neq \bar{E}$ while for $E=\bar{E}$ we put $\Omega_{\bar{E}}=1$.

The DESD can be studied using a similar approach previously outlined for DTS.
 In particular, the phases described by the topological spinor $\bm\Psi$ can be uniquely decomposed into 
 \bea
    \bm\Psi=\bm\Psi^{(G)}+\bm\Psi^{(H)}
 \eea
 where $\bm\Psi^{(G)}\in \mbox{ker}(\boldsymbol{\mathcal{H}}-\bar{E}{\bf I})$ and $\bm\Psi^{(H)}\in \mbox{im}(\boldsymbol{\mathcal{H}}-\bar{E}{\bf I})$. Similarly, the intrinsic frequencies $\bm\Omega$ can be decomposed into 
 \bea
    \bm\Omega=\bm\Omega^{(G)}+\bm\Omega^{(H)}
 \eea
 where $\bm\Omega^{(G)}\in \mbox{ker}(\boldsymbol{\mathcal{H}}-\bar{E}{\bf I})$ and $\bm\Omega^{(H)}\in \mbox{im}(\boldsymbol{\mathcal{H}}-\bar{E}{\bf I})$.
 From the dynamical equations of DESD given by Eq.(\ref{DESD}), it follows immediately that the component $\bm\Psi^{(G)}$, aligned with the ground state of $\mathcal{F}$, obeys a free, uncoupled dynamics 
\bea
    \frac{d\bm\Psi^{(G)}}{dt}=\bm\Omega^{(G)}.
    \label{PsiG}
\eea
Let us consider the expansion of  $\bm\Psi$  over the eigenstates of the Topological Dirac Equation thus getting 
\bea
    \bm\Psi=\sum_{E}c_{E}(t)\bm\Psi^{(E)},
\eea
with $c_E(t)=\bm\Psi^{\top}\bm \Psi^{({E})}$. Equation (\ref{PsiG}) can be expressed now as 
\bea
    \frac{dc_{\bar{E}}}{dt}=\Omega_{\bar{E}}.
    \label{cG}
\eea
Therefore, the component aligned with the ground state of $\mathcal{F}$ is uncoupled and describes the fact that the phases of the networks, aligned with the pattern extending on nodes and edges along the eigenstate $\bm\Psi^({\bar{E}})$, are free to oscillate at constant frequency $\Omega_{\bar{E}}$ for any value of the coupling constant $\sigma$. 
The dynamics aligned with the excited states of $\mathcal{F}$ is captured by $\bm\Psi^{(H)}$ given by 
\bea
    \bm\Psi^{(H)}=\sum_{E\neq \bar{E}}c_{E}(t)\bm\Psi^{(E)}.
\eea
In order to study these dynamics we can consider the spinor
\bea
    \bm\Theta=(\boldsymbol{\mathcal{H}}-\bar{E}{\bf I})\bm\Psi=(\boldsymbol{\mathcal{H}}-\bar{E}{\bf I})\bm\Psi^{(H)}
\eea
with the node component $\bm\alpha$ and the edge component $\bm\beta$ given by 
\bea
\bm\Theta=\begin{pmatrix}
        {\bm\alpha}\\
        {\bm\beta}
    \end{pmatrix}=\begin{pmatrix}
        -(\bar{E}-m){\bf I}_{N_0}&{\bf B}\\
        {\bf B}^{\top}& -(\bar{E}+m){\bf I}_{N_1}
    \end{pmatrix}\begin{pmatrix}
        {\bm\theta}\\
        {\bm\phi}
    \end{pmatrix},\nonumber
\eea
and eigenstate decomposition 
\bea
    \bm\Theta=\sum_{E\neq \bar{E}}(E-\bar{E})c_{E}(t)\bm\Psi^{(E)}.
\eea
The equation for $\bm\Theta$ reads
\bea
    \frac{d\bm\Theta}{dt}=(\boldsymbol{\mathcal{H}}-\bar{E}{\bf I})\bm\Omega-\sigma (\boldsymbol{\mathcal{H}}-\bar{E}{\bf I})^2\sin(\bm\Theta),
\eea
which is consistent with a freezing of the dynamics for $c_E(t)$ with $E\neq \bar{E}$ for sufficiently large coupling constant $\sigma$ and a sufficiently long time on any finite network. 
 \begin{figure*}[ht]
     \centering
     \includegraphics[width=1.8\columnwidth]{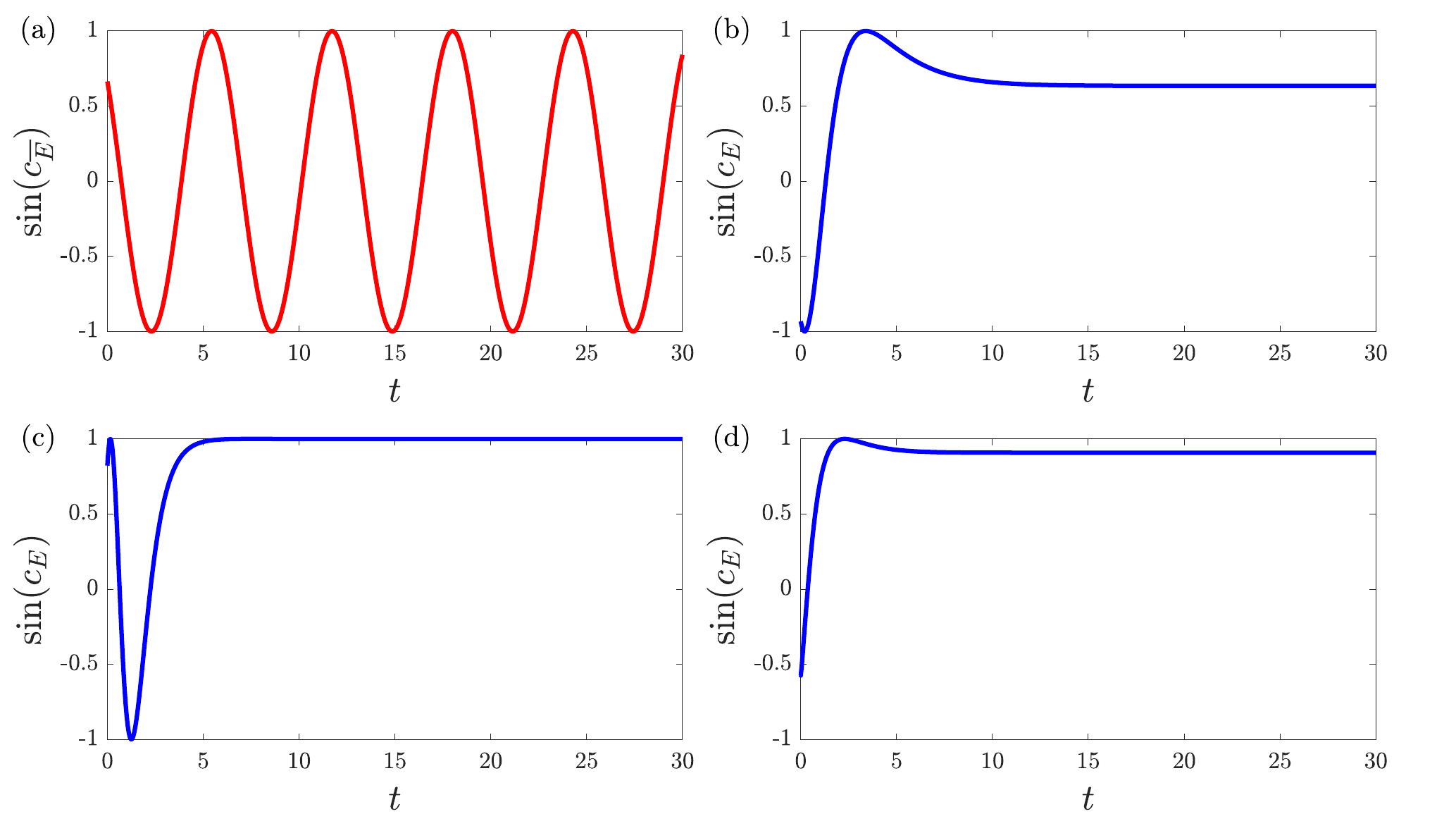}
     \caption{The DESD along the isolated (gapped) eigenstate associated with energy state $\bar{E}$ is characterized by the eigenmode that oscillates freely, while the eigenmodes corresponding to the other energy states freeze asymptotically in time.
     Panel (a) shows the time series for $\sin(c_{\bar{E}}(t))$ for the eigenmode at energy $\bar{E}=1.6642\ldots$ corresponding to the isolated Fiedler singular value $\lambda$ of the boundary operator   of a  Poisson network with $N_{0}=1000$ nodes and average degree $c=12$. Panels (b)-(d) show the time series for $\sin(c_{E}(t))$ corresponding to the three eigenmodes associated to the energies $E$ nearest to  $\bar{E}$, namely (b) $E=1.8484\ldots$, (c) $E=1.8797\ldots$, and (d) $E=1.8954\ldots$ on the same network.  The simulation of DESD is performed for a mass $m=1$ and a coupling constant $\sigma=15$.}
     \label{fig:sinc}
 \end{figure*}
We therefore conclude that for DESD the fully synchronized dynamics is encoded in the free dynamics of the component $\bm\Psi^{(G)}$ of the topological spinor aligned along the ground state of $\mathcal{F}$, and is reached when the rest of the signal leads to a frozen dynamics of $\bm\Theta$.

\begin{figure}[ht]
     \centering
     \includegraphics[width=0.8\columnwidth]{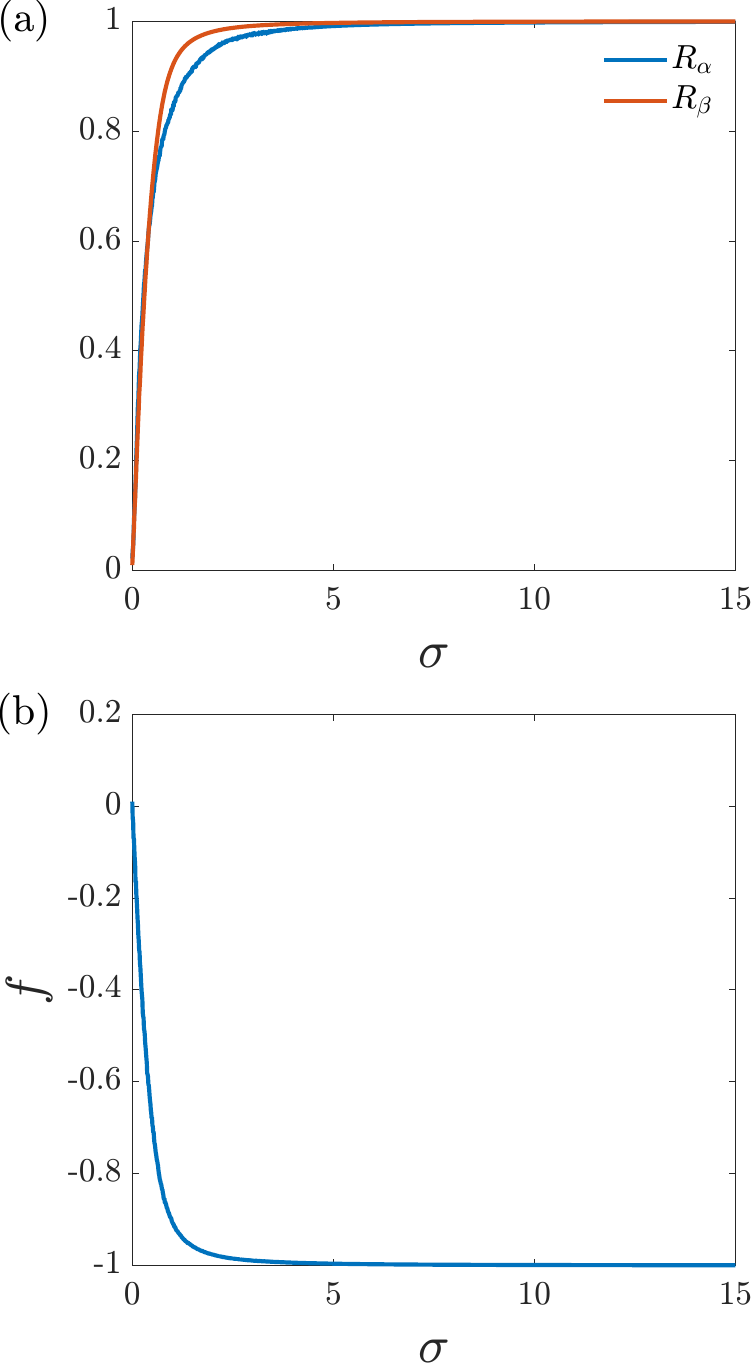}
     \caption{The existence of a topological  DESD is demonstrated by plotting in panel (a) the  order parameters $R_{\alpha}$ and $R_{\beta}$ , and in panel (b) the free energy density  $f$  as functions of the coupling constant $\sigma$.  The selected eigenstate is the Fiedler eigenstate, $\bar{E} = 1.6643$, of the same Poisson network studied in Figure \ref{fig:sinc} and the mass is $m=1$. The simulations are performed along the forward transition from random initial conditions. The equilibration time is  $T_{\text{max}} = 15$ with time steps of $dt = 0.001$. The values of $R_{\alpha}$, $R_{\beta}$, and $f$ were averaged over the final third of $T_{\text{max}}$ for each $\sigma$.}     
     \label{R_f}
 \end{figure}

In order to assess the extent to which the dynamics of $\bm\Theta$ is frozen, we can monitor the average value of the free energy $\mathcal{F}$ together with the two order parameters $R_{\alpha}$ and $R_{\beta}$ defined similarly to the DTS, i.e. 
\bea
    R_{\alpha} = \frac{1}{N_{0}}\left|\sum_{i=1}^{N_0}e^{i\alpha_{i}}\right|, \quad
    R_{\beta} = \frac{1}{N_{1}}\left|\sum_{\ell=1}^{N_1}e^{i\beta_{\ell}}\right|.
\eea
To show numerical evidence for this behavior, in Figure $\ref{fig:sinc}$ we display the time-series of $\sin(c_E(t))$ for $E=\bar{E}$ and for $E\neq \bar{E}$ on a random graph for a large value of the coupling constant $\sigma$. We observe that $\sin(c_{\bar{E}}(t))$ displays a sustained oscillation with frequency $\Omega_{\bar{E}}$ while $\sin(c_{{E}}(t))$ with $E\neq \bar{E}$ saturate to a constant value after a transient time that is longer the closer the energy $E$ is to $\bar{E}$. If we explore the DESD as a function of the coupling constant $\sigma$ (see Figure \ref{R_f}), by monitoring  the order parameters $R_{\alpha}$ and $R_{\beta}$ together with the free energy density $f=\mathcal{F}/(\sigma \mathcal{N})$ we can appreciate that the synchronized state, for a large coupling constant, is increasingly aligned to the ground state of the free energy, i.e. to the eigenstate $\bm\Psi^{(\bar{E})}$. This implies that the DESD does not occur uniformly over all nodes and edges of the networks but reflects the network patterns encoded by the eigenstate $\bm\Psi^{(\bar{E})}$ leading to a topological cluster synchronization. Indeed, in the synchronized state we have 
\bea
    {\bm\Psi}=\bm\Psi^{(G)}+\bm\Psi^{(H)}\simeq \bm\Psi^{(G)}
\eea
Thus, recalling that $\bm\Psi^{(G)}$ can be expressed as 
\bea
    \bm\Psi^{(G)}=c_{\bar{E}}(t)\bm\Psi^{(\bar{E})}
\eea
with 
\bea
    c_{\bar{E}}(t)=\Omega_{\bar{E}}t+c_{\bar{E}}(0),
\eea
and that $\bm\Psi^{(\bar{E})}$ is independent of time, then, by indicating with $\Psi_i^{(\bar{E})}$ and $\Psi_{\ell}^{(\bar{E})}$ the node and edge components of $\bm\Psi^{(\bar{E})}$ respectively, we obtain 
\bea
    \frac{d\theta_i}{dt}=\Omega_{\bar{E}}{ \Psi}^{(\bar{E})}_i,\quad \frac{d \phi_{\ell}}{dt} =\Omega_{\bar{E}}{ \Psi}^{(\bar{E})}_{\ell}
\eea
{i.e. the frequencies of the node and edge signals are proportional to the associated  components of the eigenstate $\bm\Psi^{(\bar{E})}$ }. It follows that, while in the spectral domain only the eigenmode $\bar{E}$ oscillates, in the real space domain the oscillators placed on the nodes and edges might have different frequencies, with their value  proportional to the components of the eigenstate $\bm\Psi^{(\bar{E})}$.
When the eigenstate $\bm\Psi^{(\bar{E})}$ reveals a partition of the network into clusters, DESD leads to topological cluster synchronization defined on the nodes and edges of the network.

The formulation of DESD implies the ability to design the topological cluster synchronization pattern that extends to the nodes and edges of the network by tuning the values of $\bar{E}$.
However, we note that not every eigenstate of the Topological Dirac Equation displays such a clear DESD. In particular, we observe that the most stable DESD are observed for eigenstates corresponding to gapped (or isolated) energy states $\bar{E}$. In order to investigate this matter quantitatively, in the following we study the linear stability of the DESD.

\subsection{Linear stability analysis and its numerical validation}

In this section, we will investigate the stability of DESD along the generic eigenstate ${\bm \Psi^{(E)}}$ associated with a non-degenerate energy $E$ of the Topological Dirac Equation using an approach that generalizes previous approaches used for the standard (node-based) Kuramoto model on general networks~\cite{millan2019synchronization,millan2018complex} or on lattices~\cite{hong2005collective,hong2007entrainment}.
By linearizing the DESD defined in Eq.(\ref{DESD}) for  $\|(\boldsymbol{\mathcal{H}}-\bar{E}{\bf I})\bm\Psi\|\ll 1$ we obtain 
\bea
    \frac{d\bm\Psi}{dt}=\bm\Omega-\sigma(\boldsymbol{\mathcal{H}}-\bar{E}{\bf I})^2\bm\Psi.
    \label{LDESD}
\eea
Let us expand $\bm\Psi$ and $\bm\Omega$ over the eigenstates of the Topological Dirac Equation, thus getting 
\bea
    \Psi_{r}=\sum_{E}c_{E}(t)\Psi^{(E)}_{r}\nonumber \\
    \Omega_{r}=\sum_{E}\Omega_{E}(t)\Psi^{(E)}_{r}
\eea
where $c_E(t)=\bm\Psi^{\top}\bm \Psi^{({E})}$ and $\Omega_E=\bm\Omega^{\top}\bm \Psi^{({E})} $.
Using Eq.(\ref{LDESD}) it can be readily shown that the amplitude $c_E(t)$ obeys the dynamical equation 
\bea
    \dot{c}_E(t)=\Omega_E-\sigma (E-\bar{E})^2c_E(t).
\eea
It then follows that $c_E(t)$ is given by 
\bea
    c_E(t)&=&\frac{\Omega_E}{\sigma (E-\bar{E})^2}\left(1-e^{-\sigma (E-\bar{E})^2t}\right)\nonumber \\&&+c_E(0)e^{-\sigma (E-\bar{E})^2t}
\label{cE}
\eea
for $E\neq \bar{E}$, while for $E=\bar{E}$ the solution reads
\bea
    c_{\bar{E}}(t)=\Omega_{\bar{E}}t+c_{\bar{E}}(0).
\eea
Thus, in this linear approximation, the eigenstate $E=\bar{E}$ is associated with a network pattern $\bm\Psi^{(\bar{E})}$ that continues to drift unperturbed even at very large times, while the other modes corresponding to energies $E\neq \bar{E}$ are associated with an amplitude $c_E(t)$ that, for $t\to\infty$, saturate to a fixed value $\Omega_E/(\sigma(E-\bar{E})^2)$.
Note, however, that this contribution diverges as $E\to \bar{E}$, thus if the spectral density of the eigenstates $\rho(E)$ is dense around $\bar{E}$, there is a possibility that the cumulative contributions to the phases due to the components associated with energies $E\neq \bar{E}$ diverge. When this happens, the dynamical synchronized state of DESD might not be stable in the thermodynamic limit, a feature that impedes its observability in real scenarios.

In order to study the contributions due to the component aligned with the eigenstates $E\neq \bar{E}$, we screen out the component aligned with the eigenstate associated with energy $\bar{E}$ and consider
\bea
    \bm\Phi={\bm\Psi}-\left(\bm\Psi^{\top}\bm \Psi^{(\bar{E})}\right) \bm\Psi^{(\bar{E})}.
    \label{PhiA}
\eea 
The dynamical properties of the  linearized DESD can be investigated by evaluating: (i)  the ``roughness"  $W^2$ associated with the phases $\bm\Psi$ and defined as 
\bea
    W^2=\Avg{\|\bm\Phi\|^2}=\frac{1}{\mathcal{N}}\Avg{\sum_{r=1}^\mathcal{N}\Phi_{r}^2},
    \label{W2A}
\eea
and (ii) the variance $V^2$ of the velocity of $\bm\Phi$
\bea
    V^{2}=\Avg{\|\bm\dot{\bm\Phi}\|^2}=\frac{1}{\mathcal{N}} \Avg{\sum_{r=1}^{\mathcal{N}}{{\dot{\Phi}}}_{r}^{2}}.
    \label{V2A}
\eea
A diverging value of $W^2$ observed in the thermodynamic limit $\mathcal{N}\to\infty$ will indicate that the synchronized dynamics of DESD is unstable, while a value  of $V^2$ converging to zero in the same limit will indicate that the phases are entrained.

In this work, we derive some important results that allow us to predict whether we are able to observe a stable synchronization state of the DESD, or not, depending on the properties of the density of states of the Topological Dirac Equation. These results are briefly summarized here. For details on the derivations, we refer the reader to the Appendices.
\begin{figure*}[ht]
     \centering
     \includegraphics[width=1.8\columnwidth]{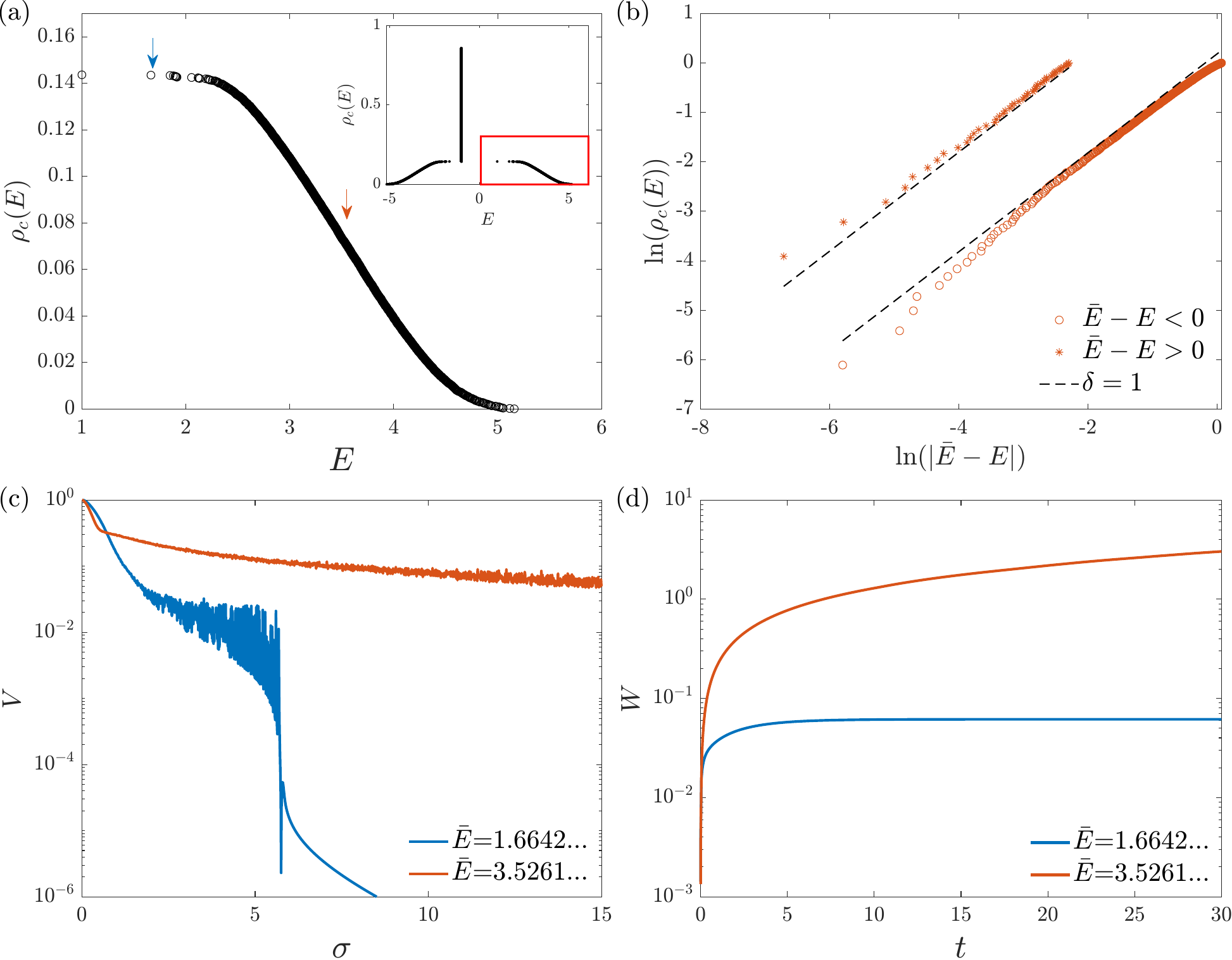}
     \caption{The stability of DESD aligned with different eigenstates $\bar{E}$ of the Topological Dirac Equation is studied numerically and compared to the predictions of the linear stability analysis. In panel (a) the positive part of the cumulative density of eigenstates $\rho_c(E)$ of the Topological Dirac Equation (with $m=1$), defined as in Figure 2, is plotted for a Poisson network with $N_0=1000$ nodes and average degree $c=12$ (the Inset shows the entire spectrum). From the positive eigenstates, we select two positive eigenstates: the  eigenstate at $\bar{E}=1.6642\ldots$ corresponding to the gapped (isolated) Fiedler eigenstate and $\bar{E}=3.5261\ldots$ located in the bulk of the spectrum. Panel (b) demonstrates that for the eigenstate in the bulk $\delta_{-}<1$. In these conditions the linear stability analysis predicts that the Fiedler eigenstate $\bar{E}=1.6642\ldots$ corresponds to a stable topological synchronized pattern while the eigenstate at energy $\bar{E}=3.5261\ldots$ is not stable. Panels (c) and (d) display the standard deviation of the velocities of the phases $V$ versus $\sigma$ and the standard deviation of the roughness of the phases $W$ versus $t$ for $\sigma=10$, showing that for the Fiedler eigenstate the DESD is stable while for the eigenstate at energy  $\bar{E}=3.5261\ldots$ DESD is not stable, in agreement with the linear stability analysis predictions.}
     \label{fig:linear}
 \end{figure*}
 
 \begin{figure*}[!htb!]
     \centering
     \includegraphics[width=1.8\columnwidth]{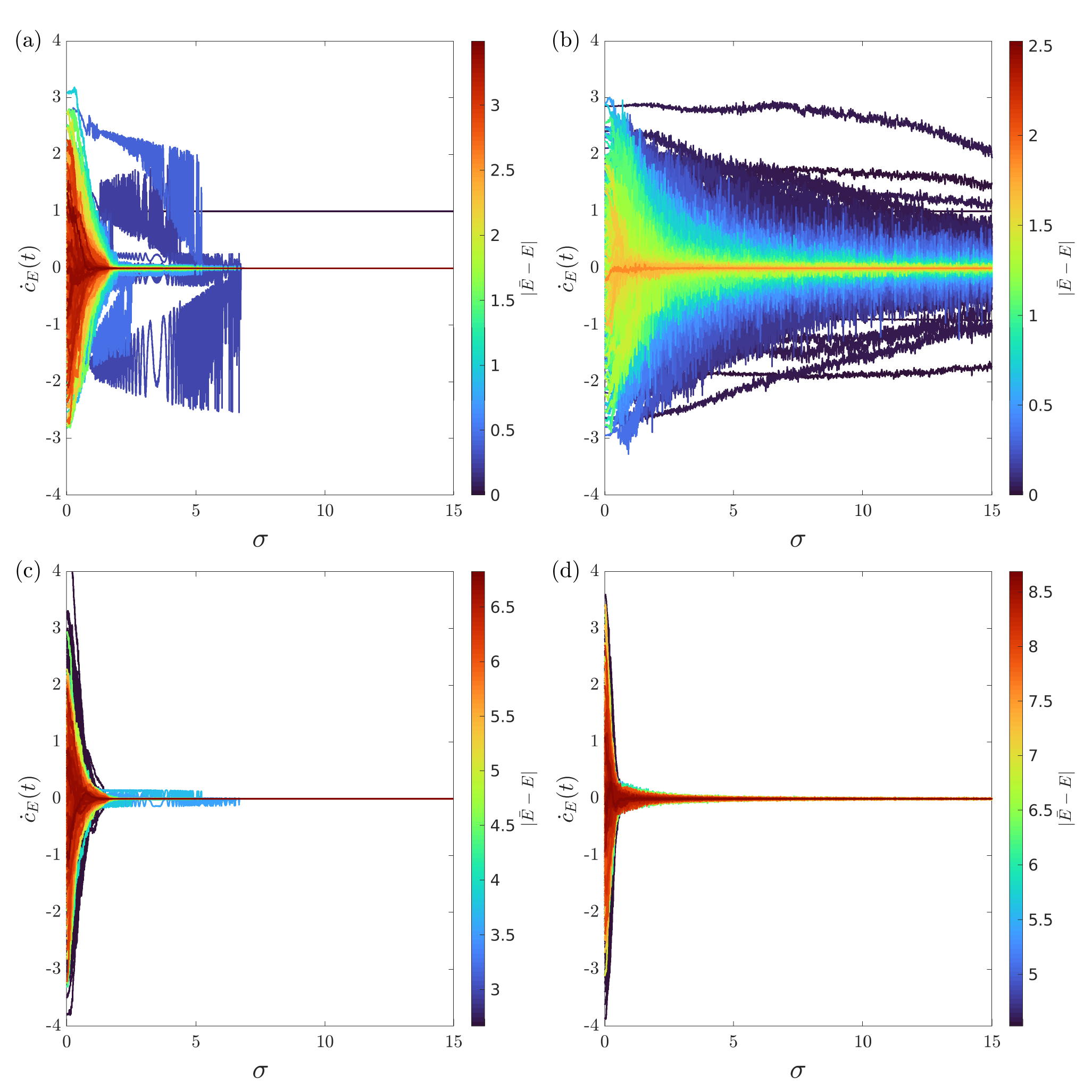}
     \caption{The stability of the DESD designed along the two eigenstates of the Poisson network considered in Figure $\ref{fig:linear}$ is studied starting from random initial conditions.
     Panels (a)--(d) show $\dot{c}_{E}(t)$ as a function of $\sigma$ for the isolated Fiedler eigenstate at energy $\bar{E}=1.6642\ldots$ (panels (a) and (c)) leading to  stable topological cluster synchronization  and for the eigenstate corresponding to the energy state $\bar{E}=3.5261\ldots$ in the bulk of the spectrum (panels (b) and (d)). The top row shows $\dot{c}_{E}(t)$ for $E>0$ and the bottom row shows $\dot{c}_{E}(t)$ for $E<0$. The color map represents the distance $|\bar{E}-E|$. It can be seen that as $\sigma$ increases, the $\dot{c}_{E}(t)$ of more eigenstates $E\neq\bar{E}$ converge to zero, with those furthest away from $\bar{E}$ being quickest (including all with $E<0$). In both cases, $\dot{c}_{{E}}$ for $E=\bar{E}$ remains constant and independent of $\sigma$, i.e. $\dot{c}_{{E}}=\Omega_{\bar{E}}=1$. Simulations are performed across the forward transitions with $\sigma$ increased adiabatically from $1$ to $15$ in steps of $0.01$. At each value of $\sigma$, the dynamics were allowed to equilibrate for $T_{\text{max}} = 15$ with time steps of $dt = 0.001$. The value of $\dot{\bm{\Psi}}(t)$, from which $\dot{c}_{E}(t)$ was calculated, was averaged over the final third of $T_{\text{max}}$ for each $\sigma$.}
     \label{fig:cdot}
 \end{figure*}


In order to characterize the properties of the density of states of the Topological Dirac Equation for $E\simeq \bar{E}$, 
we define the right spectral gap $\Delta_+$ and the left spectral gap $\Delta_-$ as 
\bea
    \Delta_+=\min_{E>\bar{E}}(E-\bar{E}),\nonumber \\
    \Delta_-=\min_{E<\bar{E}}(\bar{E}-E).
    \label{Delta}
\eea
Thus, if $\lim_{\mathcal{N}\to \infty}\Delta_{\pm}=0$ for either choice of $\pm$, or both, then the eigenstate sits in the bulk of the density of states of the Topological Dirac Equation, otherwise the eigenstate is isolated (or gapped).
As we shall discuss in detail in the following, if the eigenstate is isolated, the corresponding synchronization pattern of DESD is stable. Our criteria for assessing the stability of a synchronized state along the eigenstate $\bar{E}$ depend on the value of the exponents $\delta_{\pm}$. The exponent $\delta_+$ is defined as follows:
\begin{itemize}
    \item If $\lim_{\mathcal{N}\to \infty}\Delta_+=0$ we assume that the density of eigenstates scales as \bea\rho(E)\propto (E-\bar{E})^{\delta_+}\eea for $0<E-\bar{E}\ll 1$. 
    \item If $\lim_{\mathcal{N}\to \infty}\Delta_+>0$, we put by definition $\delta_+=\infty$.
\end{itemize}
The exponent $\delta_-$ is defined according to the following similar classification:
\begin{itemize}
    \item If $\lim_{\mathcal{N}\to \infty}\Delta_-=0$, then we assume that the density of eigenstates scales as \bea\rho(E)\propto (\bar{E}-E)^{\delta_-}\eea for $0<\bar{E}-E\ll 1$.
    \item  If $\lim_{\mathcal{N}\to \infty}\Delta_->0$, we put by definition $\delta_-=\infty.$
\end{itemize}
 
We are now in a position to summarize the results (see Appendices) obtained by the linear stability approach for the synchronized state of DESD in terms of the values of $\delta_\pm$:
\begin{itemize}
    \item If both $\delta_+=\infty$ and $\delta_-=\infty$, the synchronized state of DESD is stable. In this case, we will have $W^2\to\mbox{cost}$  and $V^2\to 0$ as $\mathcal{N}\to \infty$.
    \item If either $\delta_+\leq3$ or $\delta_-\leq3$, and if they are both greater than one, the phases are entrained, but the synchronized state of DESD is not thermodynamically stable. In this case, we will have $W^2\to \infty$ and $V^2\to 0$ as $\mathcal{N}\to \infty$.
    \item If either $\delta_+\leq1$ or $\delta_-\leq1$, then the entrained state might not even exist. In this case, we will have $W^2\to \infty$ and $V^2$ might not converge to zero as $\mathcal{N}\to \infty$ since the linear approximation fails.
\end{itemize}
In order to numerically validate these theoretical predictions, we considered DESD defined on a random network of $N_{0}=1000$ and average degree $c=12$. By setting $m=1$, we first numerically evaluated the cumulative density of states $\rho_c(E)$ (see panel (a) of Figure \ref{fig:linear}). We have focused on two specific eigenstates of positive energy $E$. The first eigenstate of energy $\bar{E}=1.6642\ldots$ corresponds to the Fiedler singular state of the boundary operator, and is isolated from the rest of the spectrum, thus $\delta_{\pm}=\infty$. The second eigenstate of energy $\bar{E}=3.5261\ldots$ is deep in the bulk of the spectrum and the density of states is characterized by $\delta_{-}<1$ (see panel (b) of Figure \ref{fig:linear}).  
We study the stability of the DESD by considering the full nonlinear DESD starting from an initial condition that is as small perturbation to the topological synchronized state.
In particular, we take
\bea
    \bm\Psi(0)=\bm\Psi^{(\bar{E})}+\epsilon\frac{{\bf x}}{|{\bf x}|}
\eea
where ${x}_r\sim \mathcal{N}({0,1})$ and $\epsilon=0.1$. 
In panel (c) and (d) of Figure \ref{fig:linear} we plot $V$ as a function of $\sigma$ and $W$ as a function of time $t$ for $\sigma=10$ for the DESD designed along the eigenstates $\bar{E}=1.6642\ldots$ and $\bar{E}=3.5261\ldots$. Our results show that, consistently with our theory, for the first considered eigenstate, the phases are entrained for large values of $\sigma$, i.e., $V$ is very small in the large time limit, and the synchronized state admits small fluctuations, corresponding to a small roughness $W$. However, for the second eigenstate we observe large fluctuations $W$ around the topological synchronized state, which increase over time, indicating that the synchronized state of DESD is not stable for this choice of the eigenstate $\bar{E}$. 

In Figure \ref{fig:cdot}, we provide further numerical evidence of the stability, or lack thereof, of DESD aligned with the two mentioned eigenstates of the Topological Dirac Equation corresponding to the Fiedler eigenstate $\bar{E}=1.6642\ldots$ (stable) and the eigenstate in the bulk of the spectrum $\bar{E}=3.5261\ldots$ (unstable).
Starting from random initial conditions, we plot $\dot{c}_{E}(t)$ for positive values of the energy (panels (a)-(b)) and negative values of the energy (panels (c)-(d)). We show that for both cases the eigenmode at energy $E=\bar{E}$ oscillates unperturbed with frequency $\dot{c}_{\bar{E}}=\Omega_{\bar{E}}=1$ while the negative energy eigenmodes freeze rapidly by increasing the coupling constant $\sigma$. However, we observe a sharp difference in the dynamical behavior of positive energies with $E\simeq \bar{E}$ and $E\neq \bar{E}$.
Indeed we have that for the stable topological DESD (panel (a)) $\dot{c}_E$ freezes quickly with $\sigma$ albeit eigenmodes closer to $\bar{E}$ are the last to freeze, while for the unstable topological DESD the freezing is much slower and is to be understood as a finite size effect.

\begin{figure}[!htb!]
     \centering
     \includegraphics[width=0.8\columnwidth]{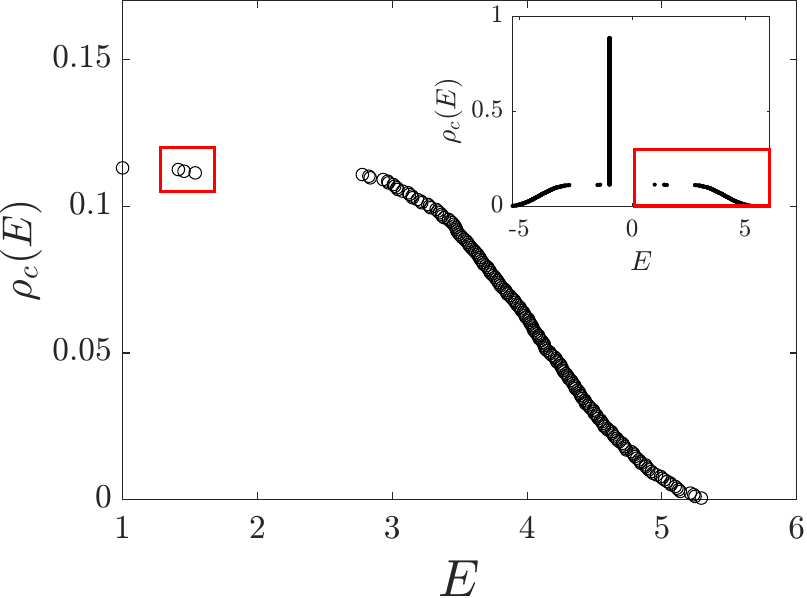}
     \caption{The positive part of the cumulative density of eigenstates $\rho_c(E)$ of the Topological Dirac Equation (with $m=1$), defined as in Figure \ref{spectra}, is plotted for a SBM with $4$ clusters, $50$ nodes per cluster, an intra-cluster connection probability of $p_1 = 0.3$, yielding an average intra-cluster degree of $c_1 = 15$, and an inter-cluster connection probability of $p_2 = 0.007$, corresponding to an average inter-cluster degree of $c_2 = 1$ (the Inset shows the entire spectrum). The red box in the main figure highlights the energies associated with eigenstates that partition the SBM. These are $E=1.4137\ldots$, $1.4557\ldots$, and $1.5388\ldots$ and correspond to the lowest energies that satisfy $E\neq m$.}
     \label{fig:SBM_spec}
 \end{figure}

\subsection{Designing stable topological cluster synchronization patterns on the stochastic block model (SBM)}

\begin{figure*}[htb]
     \centering
     \includegraphics[width=\linewidth]{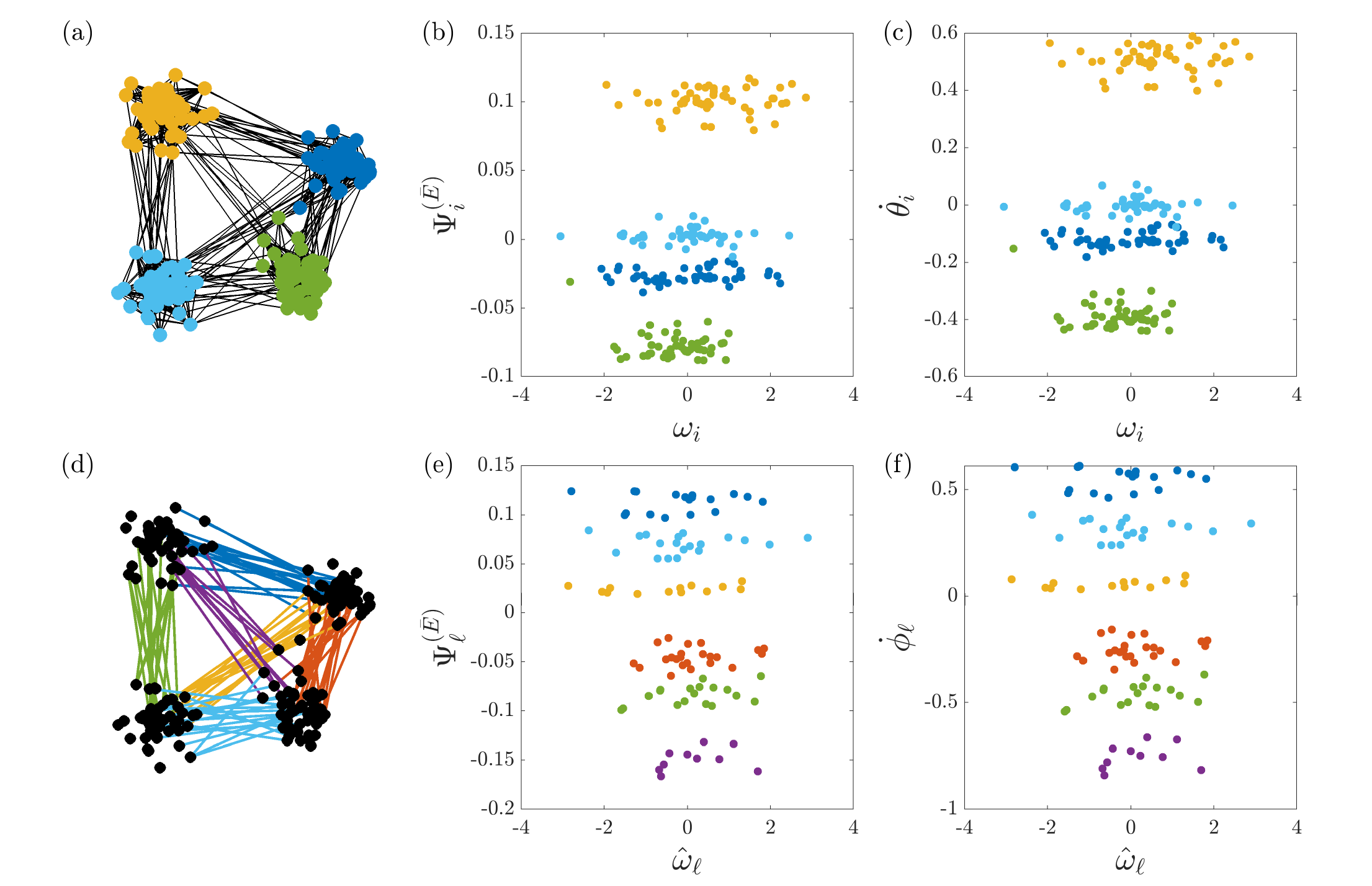}
     \caption{ The DESD leads to topological cluster synchronization defined on nodes and edges whose frequencies settle to values proportional to the elements of the eigenstate $\bm\Psi^{\bar{E}}$.
     The considered stochastic block model (SBM) is visualized in panel (a) with each cluster highlighted in a unique color.  Panel (b) shows the element $\Psi_{i}^{(\bar{E})}$ of the eigenstate associated  on the generic node $i$ of the network which determine  the frequency of $\dot{\theta}_i$ of the node (panel c) in the DESD regardless of the values of the node intrinsic frequency $\omega_i$. 
     Panel (d) provides a visualization of  the different bundles of edges connecting the different communities in the same network  indicated with different colors (the edges among nodes of the same community are omitted).
     Panels (e) and (f) compare the elements $\Psi_{\ell}^{(\bar{E})}$ associated with the generic edge $\ell$ of the eigenstate  with the frequencies $\dot{\phi}_{\ell}$ of the phases associated with it in the DESD regardless to its intrinsic frequency $\hat{\omega}_i$. The DESD is run up to a maximum time $T_{\text{max}}=30$  at $\sigma=15$.} 
     \label{fig:SBM_comp}
 \end{figure*}

\begin{figure*}[!ht!]
     \centering
     \includegraphics[width=\linewidth]{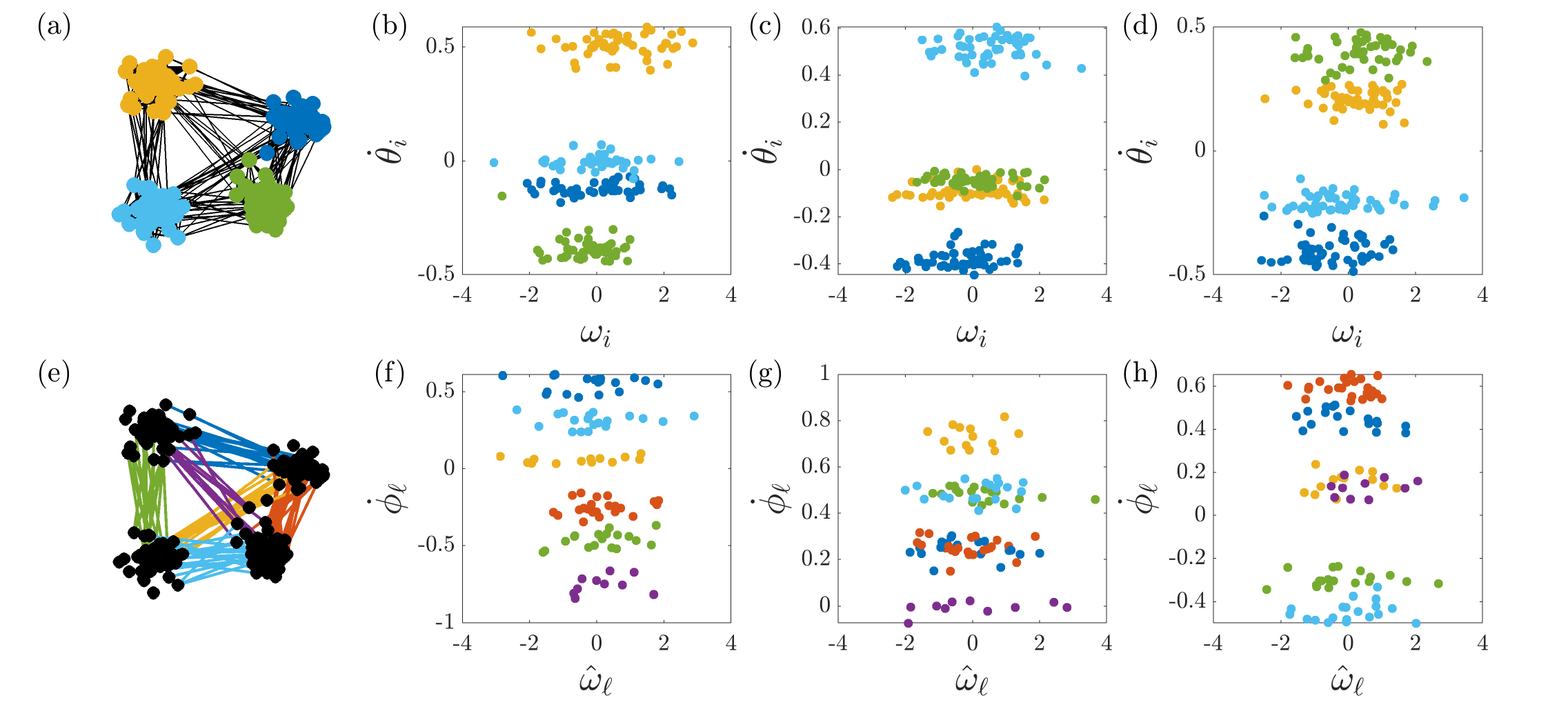}
     \caption{The DESD allows us to design topological cluster synchronization pattern defined on nodes and edges on the SBM.
     By considering the DESD aligned with the isolated eigenstates of the Topological Dirac Equation of the SBM shown in Figure $\ref{fig:SBM_spec}$ we show that DESD leads to cluster synchronization of nodes and edges where different communities and different bundles of links are partition into clusters that oscillate at the same frequency independently of their intrinsic frequency. Here panel (a) and (e) define the color code of the node clusters and the edge clusters while panels 
     panels (b)–-(d)  show the phase velocities of the nodes, $\dot{\bm\theta}$, and the edges, $\dot{\bm\phi}$, plotted against their intrinsic frequencies, $\bm\omega$ and $\bm\hat{\omega}$, for the DESD associate the  eigenstates $\bar{E}=\pm 1.4137\ldots$ ((b) and (f)), $\bar{E}=\pm 1.4557\ldots$ ((c) and (g)), and $\bar{E}=\pm 1.5388\ldots$ ((d) and (h)). The DESD is run up to a maximum time $T_{\text{max}}=30$  at $\sigma=15$. }
     \label{fig:SBM_dyn}
 \end{figure*}

The SBM provides an ideal setting for demonstrating the design of topological cluster synchronization patterns via the DESD model. The Dirac spectrum of the SBM (see Figure $\ref{fig:SBM_spec}$) exhibits isolated eigenstates at energies slightly above $E=m$ or slightly below $E=-m$. Their associated eigenstates have a node and an edge sectors that naturally partition the graph into node clusters and edge bundles. 
Indeed, these eigenstates have the structure given by Eq.(\ref{eq:Dirac_eigestates}) where ${\bf u}_\lambda$ and ${\bf v}_\lambda$ are singular vectors of the boundary operator, i.e. eigenvectors of the graph Laplacian ${\bf L}_{[0]}$ and the ${\bf L}_{[1]}$ Hodge Laplacian often used in spectral clustering ~\cite{capocci2005detecting,von2007tutorial}. The relative normalization of the node component and the edge component of the Dirac eigenstate is set by the non-zero value of the mass $m$. In particular, for $m>0$ the positive eigenstates have a more significant contribution associated with the nodes while the negative eigenstates have a more significant contribution associated with the edges.

In DESD, by setting $\bar{E}$ to the energy of any of the isolated eigenstates, the synchronization dynamics inherit the eigenstate’s partitioning properties. Indeed, since the DESD occurs along the eigenstate $\bm\Psi^{\bar{E}}$, DESD leads to topological cluster synchronization. In this dynamical state, nodes belonging to different communities and edges belonging to different bundles oscillate in the real space at different frequencies with the frequencies proportional to the value of the eigenstate $\bm\Psi^{\bar{E}}$. This effect is illustrated in Figure~\ref{fig:SBM_comp}. Panels (a) and (d) highlight the node clusters and inter-cluster edge bundles, respectively. Panels (b) and (e) show the components of the eigenstate associated with the Fiedler energy ($\bar{E} = 1.4137\dots$) on the nodes ($\Psi_{i}^{(\bar{E})}$) and on the inter-cluster edges ($\Psi_{\ell}^{(\bar{E})}$). For direct comparison with panels (c) and (f), these components of the eigenstate are plotted against the intrinsic frequencies of the nodes and edges, $\omega$ and $\hat{\omega}$, respectively. Panels (c) and (f) then show the resulting phase velocities of the node ($\dot{\theta}$) and edge ($\dot{\phi}$) signals, again as functions of $\omega$ and $\hat{\omega}$. A direct comparison reveals that the phase velocities synchronize with the Fiedler eigenstate, independently of the intrinsic frequencies.

The eigenstates associated with the negative counterparts of the isolated positive energies exhibit identical structural partitions, differing only in the relative scaling of the node and edge components. Specifically, node signals are amplified relative to edge signals for positive energies, and vice versa for negative ones. Thus, both positive and negative Fiedler eigenstates induce the same partitioning, though only the positive eigenstate is used in panels (b) and (c), and its negative counterpart in panels (e) and (f).

A notable property of DESD on the SBM is the possibility to design several topological cluster synchronization patterns on the same network thanks to the presence of several isolated eigenstates of the Dirac spectrum (see Figure \ref{fig:SBM_dyn}).
In order to provide evidence for this remarkable property of DESD, we consider the three isolated Dirac eigenstates highlighted on the spectrum of the SBM shown in Figure $\ref{fig:SBM_spec}$.
The DESD allows us to design different topological cluster synchronization patterns by selecting any of these different eigenstates. The choice of the eigenstate indeed produces a topological cluster synchronization on nodes and edges that partition both nodes and edges into node clusters and edge bundles oscillating at the same frequency independently of their intrinsic frequency.
As we have shown in Sec.\ref{preamble}, this phenomenology of SBM is also shared by real networks with modular structure, such as the connectome network shown in Figure $\ref{fig:brain_network}$.


\section{Conclusion}
In this work, we proposed DESD which allows us to design topological cluster synchronization patterns extending to node and edge signals of a network encoded in the topological spinor.
These topological cluster synchronization patterns are such that oscillators associated with different clusters of nodes and bundles of edges oscillate at distinct frequencies. These frequencies are dictated by the elements of the eigenstate corresponding to the fundamental state of the free energy associated with the DESD.
The design principle adopted in DESD modulates this ground state with a procedure similar to the one adopted in optimization algorithms used in statistical mechanics approaches such as simulated or quantum annealing.
Here, the fundamental state of the free energy is taken to be one of the eigenstates of the Topological Dirac Equation which provides a very efficient way to decompose the dynamical state of topological spinors.
We provide evidence that topological cluster synchronization states can be achieved on  random graphs, stochastic block models and real network datasets provided that their associated eigenstate is gapped (isolated). Moreover, if this condition is not met, we provide conditions for the stability of the DESD defined along an eigenstate in the bulk of the Dirac spectrum of the network.
On random graphs, we can design stable topological cluster synchronization very reliably along the Fiedler eigenstate of the network. On the stochastic block model, DESD can induce different stable topological cluster synchronization states on nodes and edges which correlate with different partitions of the blocks and the edges connecting them. Similarly, on the considered connectome dataset, DESD can provide reliable partitions of the nodes and edges of the network revealing the  left–right hemispheric separation and an anterior–posterior differentiation.

This work can be extended in different directions. From a theoretical perspective, an open question is to investigate topological cluster synchronization on higher-order networks where topological signals are not only defined on nodes and edges, but also on triangles, tetrahedra and so on. On the more applied side, further exploration of the design principle on real modular networks, such as connectomes, might shed light in the interplay between network structure and function.

\newpage


\appendix

\section{Appendix A: Stability of DESD}
In this Appendix, we will investigate the stability of DESD along the generic eigenstate ${\bm \Psi^{(E)}}$ associated with an energy $\bar{E}$ of the Topological Dirac Equation. 
To this end, we follow the approach used for the standard Kuramoto model on simple networks and on lattices \cite{millan2019synchronization,millan2018complex,hong2005collective,hong2007entrainment} and define $\bm\Phi$ as the contribution to the phases $\bm\Psi$ that is orthogonal to the eigenstate $\bm\Psi^{(\bar{E})}$, i.e.  
\bea
\bm\Phi={\bm\Psi}-\left(\bm\Psi^{\top}\bm \Psi^{(\bar{E})}\right) \bm\Psi^{(\bar{E})}.
\eea 
In order to study the stability of the DESD we investigate the ``roughness"  $W^2$ associated with the phases $\bm\Psi$ calculated for the linearized DESD (Eq.(\ref{LDESD})) and defined as 
\bea
W^2=\Avg{\|\bm\Phi\|^2}=\frac{1}{\mathcal{N}}\Avg{\sum_{r=1}^\mathcal{N}\Phi_{r}^2}.
\label{W2}
\eea
Using   the orthogonality of the eigenstates of the Topological Dirac Equation we thus obtain
\bea
W^2=\frac{1}{\mathcal{N}}\sum_{E\neq \bar{E}}\Avg{c_{E}^{2}(t)}.
\label{W2c}
\eea
The explicit expression of $c_E(t)$ is given by Eq.(\ref{cE}). Explicitly performing the average over the intrinsic frequencies $\Omega_E$, we obtain 
\begin{widetext}
\bea
    \Avg{c_{E}^{2}(t)} &=& \Avg{\left [\frac{\Omega_E}{\sigma (E-\bar{E})^2}\left(1-e^{-\sigma (E-\bar{E})^2t}\right)+c_E(0)e^{-\sigma (E-\bar{E})^2t}\right ]^2} \nonumber \\&=& \frac{1}{\sigma^2(E-\bar{E})^{4}}(1-e^{-\sigma (E-\bar{E})^2t})^{2} + c_{E}^{2}(0)e^{-2\sigma (E-\bar{E})^2t},\eea
    \end{widetext}
    where we have assumed $\Avg{\Omega_{E}} = 0$ and $\Avg{\Omega_{E}\Omega_{E^{\prime}}}=\delta_{E,E^{\prime}}$. 
    Thus, inserting this expression in Eq.(\ref{W2c}) we get, in the limit $t\to \infty$,  the expression for $W^2$ given by
\bea
    W^{2}= W^2_{+}+W^2_{-}\eea
    with \bea
    W^2_+&=&\frac{1}{\mathcal{N}}\sum_{E> \bar{E}}\frac{1}{\sigma^2 (E-\bar{E})^4},\nonumber \\
    W^2_-&=&\frac{1}{\mathcal{N}}\sum_{E< \bar{E}}\frac{1}{\sigma^2 (E-\bar{E})^4}
\eea
It is therefore sufficient that either $W^2_+$ or $W^2_-$ diverges in the thermodynamic limit, i.e., for $\mathcal{N}\to \infty$, for the synchronization state of DESD to lose its stability.
Let us now determine the spectral properties of the Topological Dirac Equation that will determine whether $W^2_{+}$ diverges in the thermodynamic limit. A similar argument will also hold for $W^2_-$.
Let us define the right spectral gap $\Delta_+$ and the left spectral gap $\Delta_-$ associated with the selected energy state $\bar{E}$ of the Dirac operator 
\bea
\Delta_+=\min_{E>\bar{E}}(E-\bar{E}),\nonumber \\
\Delta_-=\min_{E<\bar{E}}(\bar{E}-E).
\label{Delta_app}
\eea
In this scenario, for  $\mathcal{N}\ll 1$ we can express $W^2_+$ and $W^2_-$ in the continuum approximation as 
\bea
W^2_+=\int_{E>\bar{E}+\Delta_+} dE\frac{ \rho(E)}{\sigma^2 (E-\bar{E})^4},\nonumber \\
W^2_-=\int_{E<\bar{E}-\Delta_-} dE \frac{ \rho(E)}{\sigma^2 (E-\bar{E})^4}.
\eea
Thus, we have different scenarios for $W^2_+$ in the limit $\mathcal{N}\to \infty$:
\begin{itemize}\item 
If $\lim_{\mathcal{N}\to \infty}\Delta_+>0$, then $W^2_+$ is finite.
\item 
If $\lim_{\mathcal{N}\to \infty}\Delta_+=0$, then, assuming that the density of eigenstates scales as $\rho(E)\propto (E-\bar{E})^{\delta_+}$ for $0<E-\bar{E}\ll 1$, we obtain for any finite $\mathcal{N}$, 
\bea
 W^{2}_+ \sim
    \begin{cases}
        \Delta_+^{-(3-\delta_+)} & \text{if $\delta_+<3$,}\\
        -\ln(\Delta_+) & \text{if $\delta_+=3$,}\\
        \text{finite} & \text{if $\delta_+>3$},\\
    \end{cases}
    \eea
    as long as $0<E-\bar{E}\ll 1$.
    Thus, as $\mathcal{N}\to \infty$, $\Delta_+\to 0$, $W^2_+$ diverges for $\delta_+\leq 3$ and the synchronized patterns of the DESD cannot be thermodynamically stable.
\end{itemize}

Similar scenarios hold for $W^2_-$:
\begin{itemize}\item 
If $\lim_{\mathcal{N}\to \infty}\Delta_->0$, then $W^2_-$ is finite.
\item 
If $\lim_{\mathcal{N}\to \infty}\Delta_-=0$ then, assuming that the density of eigenstates scales as $\rho(E)\propto (\bar{E}-E)^{\delta_-}$ for $0<\bar{E}-E\ll 1$, we obtain for any finite $\mathcal{N}$,
\bea
 W^{2}_- \sim
    \begin{cases}
        \Delta_-^{-(3-\delta_-)} & \text{if $\delta_-<3$,}\\
        -\ln(\Delta_-) & \text{if $\delta_-=3$,}\\
        \text{finite} & \text{if $\delta_->3$,}\\
    \end{cases}
    \eea
     as long as $0<\bar{E}-E\ll 1$.
     Thus, as $\mathcal{N}\to \infty$, $\Delta_-\to 0$, $W^2_-$ diverges for $\delta_-\leq 3$ and the synchronized patterns of the DESD cannot be thermodynamically stable.
\end{itemize}

\section{Appendix B: Validity of the linear approximation}
The linear approximation of the DESD given by Eq. (\ref{LDESD}) is valid when the elements of the argument of the sine function are small, i.e. $(\boldsymbol{\mathcal{H}}-\bar{E}{\bf I})\bm\Psi$ is small. A useful global parameter for establishing the presence or absence of this condition is the correlation $C$ defined as follows:
\bea
    C = \Avg{\|(\boldsymbol{\mathcal{H}}-\bar{E}{\bf I})\bm\Psi\|^2}=\frac{1}{\mathcal{N}}{\bm \Psi}^{\top}[\boldsymbol{\mathcal{H}}-\bar{E}{\bf I}]^{2}{\bm \Psi},
\eea
where we have used the fact that ${\boldsymbol{\mathcal{H}}}-\bar{E}{\bf I}$ is a symmetric matrix. The divergence of $C$ in the large network limit establishes the failure of the linear approximation. Using the orthogonality of the eigenstates, the correlation $C$ can be expressed as 
\begin{equation} \label{corrproj}
    C = \frac{1}{\mathcal{N}}\Avg{\sum_{E\neq \bar{E}}c_{E}^{2}(t)(E-\bar{E})^{2})}
\end{equation}
By substituting Eq. (\ref{cE}) for $c_{E}(t)$ for $E\neq \bar{E}$ into Eq. (\ref{corrproj}) we obtain
\begin{widetext}
\begin{equation}
    C = \frac{1}{\mathcal{N}}\sum_{E\neq \bar{E}}\left[\left(\frac{(1-e^{-\sigma (E-\bar{E})^2t})^2}{\sigma^{2}(E-\bar{E})^{4}}+c_{E}^{2}(0)e^{-2\sigma (E-\bar{E})^2t}\right)(E-\bar{E})^{2}\right]
\end{equation}
\end{widetext}
which gives in the asymptotic limit $t\rightarrow\infty$:
\begin{equation}
    C = \frac{1}{\mathcal{N}}\sum_{E\neq \bar{E}}\frac{1}{\sigma^{2}(E-\bar{E})^{2}}
\end{equation}
In the large network limit $\mathcal{N}\ll 1$ we have:
\bea
    C = C_++C_-\eea
    with \bea
 C_+  &=& \int_{E>\bar{E}+\Delta_+} dE\, \frac{\rho(E)}{\sigma^{2}(E-\bar{E})^2},\nonumber \\
 C_-  &= &\int_{E<\bar{E}-\Delta_-} dE\, \frac{\rho(E)}{\sigma^{2}(E-\bar{E})^2},
\eea
where $\Delta_{\pm}$ are defined in Eq.(\ref{Delta_app}).
Thus, we have different scenarios for $C_+$:
\begin{itemize}\item 
If $\lim_{\mathcal{N}\to \infty}\Delta_+>0$, then $C_+$ is finite.
\item 
If $\lim_{\mathcal{N}\to \infty}\Delta_+=0$ then, assuming that the density of eigenstates scales as $\rho(E)\propto (E-\bar{E})^{\delta_+}$ for $0<E-\bar{E}\ll 1$, we obtain for any finite $\mathcal{N}$, 
\bea
 C_+ \sim
    \begin{cases}
        \Delta_+^{-(1-\delta_+)} & \text{if $\delta_+<1$,}\\
        -\ln(\Delta_+) & \text{if $\delta_+=1$,}\\
        \text{finite} & \text{if $\delta_+>1$},\\
    \end{cases}
    \eea
    as long as $0<E-\bar{E}\ll 1$.
    Thus, as $\mathcal{N}\to \infty$, $\Delta_+\to 0$, $C_+$ diverges for $\delta_+\leq 1$ and the linearization of the DESD loses its validity.
\end{itemize}
Similar scenarios hold for $C_-$:
\begin{itemize}\item 
If $\lim_{\mathcal{N}\to \infty}\Delta_->0$, then $C_-$ is finite.
\item 
If $\lim_{\mathcal{N}\to \infty}\Delta_-=0$ then, assuming that the density of eigenstates scales as $\rho(E)\propto (\bar{E}-E)^{\delta_-}$ for $0<\bar{E}-E\ll 1$, we obtain
\bea
 C_- \sim
    \begin{cases}
        \Delta_-^{-(1-\delta_-)} & \text{if $\delta_-<1$,}\\
        -\ln(\Delta_-) & \text{if $\delta_-=1$,}\\
        \text{finite} & \text{if $\delta_->1$,}\\
    \end{cases}
    \eea
     as long as $0<\bar{E}-E\ll 1$.
Thus, as $\mathcal{N}\to \infty$, $\Delta_-\to 0$, $C_-$ diverges for $\delta_+\leq 1$ and the linearization of the DESD loses its validity.
\end{itemize}


\section{Appendix C: Entrained phases}
Here we characterise the fluctuation in ${\bm {\dot \Phi}}$ using the global parameter $V^{2}$ defined as follows:
\bea\label{V2}
    V^{2}=\Avg{\|\bm\dot{\bm\Phi}\|^2}=\frac{1}{\mathcal{N}} \Avg{\sum_{r=1}^{\mathcal{N}}{{\dot{\Phi}}}_{r}^{2}}\eea
    Using the orthogonality of the eigenstates we have 
    \bea
    V^2=  \frac{1}{\mathcal{N}}\sum_{E\neq \bar{E}}\Avg{{\dot c}_{E}^{2}(t)}
\eea
Using the linearized solution (\ref{cE}), we obtain the following expression for ${\dot c}_{E}(t)$ for $E \neq \bar{E}$:
\bea\label{cdott}
    {\dot c}_{E}(t) = \left[\Omega_{E}-\sigma (E-\bar{E})^2c_{E}(0)\right]e^{-\sigma(E-\bar{E})^2t}
\eea
Substituting Eq. (\ref{cdott}) into Eq. (\ref{V2}), performing the average, and taking the limits $t\rightarrow\infty$ and $\mathcal{N}\rightarrow\infty$ we find that  $V^{2}$ can be expressed as
\bea
    V^2\sim  \frac{1}{\mathcal{N}}\sum_{E\neq \bar{E}}[1+\sigma^2(E-\bar{E})^4c_{E}^2(0)]e^{-2\sigma(E-\bar{E})^{2}t}.
\eea
In the limit $\mathcal{N}\gg1$ we can perform the continuous approximation 
\bea
    V^{2}=V^2_++V^2_-\eea
    with 
    \bea V^2_+&=&\int_{E> \bar{E}+\Delta_+} dE\,\rho(E)[1+\sigma^2(E-\bar{E})^4c_{E}^2(0)]e^{-2\sigma(E-\bar{E})^{2}t},\nonumber \\
    V^2_-&=&\int_{E< \bar{E}-\Delta_-} dE\,\rho(E)[1+\sigma^2(E-\bar{E})^4c_{E}^2(0)]e^{-2\sigma(E-\bar{E})^{2}t}.\nonumber
\eea
To the leading term we thus have 
\bea
V^2_+&\sim &\int_{E> \bar{E}+\Delta_+} dE\,\rho(E)e^{-2\sigma(E-\bar{E})^{2}t},\nonumber \\
V^2_-&\sim&\int_{E< \bar{E}-\Delta_-} dE\,\rho(E)e^{-2\sigma(E-\bar{E})^{2}t}.
\eea
We observe that under very general conditions, as long as the linear approximation of DESD is valid, the phases become entrained, meaning that $V^2\to 0$ as $t\to \infty$ for $\mathcal{N}\gg 1$.
In order to show this, let us consider the following scenarios for $V^2_+$
\begin{itemize}
\item
If $\lim_{\mathcal{N}\to \infty}\Delta_+>0$ in the limit $\mathcal{N}\to \infty$ we obtain 
\bea
V^2_+\sim e^{-2\sigma\Delta_+^2 t}\to 0 &\ \mbox{for} \ &t\to \infty.
\eea
\item 
If $\lim_{\mathcal{N}\to \infty}\Delta_+=0$ then, assuming that the density of eigenstates scales as $\rho(E)\propto (E-\bar{E})^{\delta_+}$ for $0<E-\bar{E}\ll 1$, we obtain for any finite $\mathcal{N}$, 
\bea
V^2_+\sim t^{-2/\delta_+}\to 0 &\ \mbox{for} \ &t\to \infty.
\eea
\end{itemize}
Similarly, for $V^2_-$ we obtain 
\begin{itemize}
\item
If $\lim_{\mathcal{N}\to \infty}\Delta_->0$ in the limit $\mathcal{N}\to \infty$ we obtain 
\bea
V^2_-\sim e^{-2\sigma\Delta_-^2 t}\to 0 &\ \mbox{for} \ &t\to \infty.
\eea
\item 
If $\lim_{\mathcal{N}\to \infty}\Delta_-=0$ then, assuming that the density of eigenstates scales as $\rho(E)\propto (\bar{E}-E)^{\delta_-}$ for $0<\bar{E}-E\ll 1$, we obtain for any finite $\mathcal{N}$, 
\bea
V^2_-\sim t^{-2/\delta_-}\to 0 &\ \mbox{for} \ &t\to \infty.
\eea
\end{itemize}


\bibliographystyle{unsrt}
\bibliography{main.bib}

\end{document}